\documentclass[]{JHEP3}
\newcommand{\field}[1]{\mathbb{#1}} 
\newcommand*{\Dsl}[0]{{\rlap{\kern2.25pt /}{D}}}
\newcommand*{\Asl}[0]{{\rlap{\kern2.25pt /}{A}}}
\newcommand*{\dsl}[0]{{\rlap{\kern0.5pt /}{\partial}}}
\newcommand*{\xisl}[0]{{\rlap{\kern0.5pt /}{\xi}}}
\newcommand*{\asl}[0]{{\rlap{\kern0.5pt /}{a}}}
\newcommand*{\bsl}[0]{{\rlap{\kern0.5pt /}{b}}}

\newcommand*{\Tr}[0]{{\rm Tr}}

\usepackage{amsmath}
\usepackage{amsfonts}
\usepackage{graphicx,graphics}

\title{Phase diagrams of $SU(N)$ gauge theories with fermions in various representations}
\author{Joyce C. Myers\\ Swansea University, Physics Department, Swansea SA2 8PP, UK\\ E-mail: \email{j.c.myers@swan.ac.uk}}
\author{Michael C. Ogilvie\\ Washington University, Physics Department, St. Louis, MO 63130 USA\\ E-mail: \email{mco@wuphys.wustl.edu}}

\abstract{
We minimize the one-loop effective potential for $SU(N)$ gauge theories including fermions with finite mass in the fundamental (F), adjoint (Adj), symmetric (S), and antisymmetric (AS) representations. We calculate the phase diagram on $S^1 \times {\field R}^3$ as a function of the length of the compact dimension, $\beta$, and the fermion mass, $m$, for various $N$ and $N_f$. We consider the effect of periodic boundary conditions [PBC(+)] on fermions as well as antiperiodic boundary conditions [ABC(-)]. With standard ABC(-) on fermions only the deconfined phase is found at one-loop for all representations considered. However, the use of PBC(+) produces a rich phase structure. These phases are distinguished by the eigenvalues of the Polyakov loop $P$. In the case of fundamental representation fermions [QCD(F,+)], a phase in which ${\text Re} \, \Tr P$ is minimized (and negative) is favoured for all values of $m \beta$. For $N$ odd charge conjugation (${\cal C}$) symmetry is spontaneously broken in this phase due to ${\cal O} (1/N)$ effects. Minimization of the effective potential for QCD(AS/S,+) results in a phase where $\left| {\text Im} \, \Tr P \right|$ is maximized, resulting in ${\cal C}$-breaking for all N and all values of $m \beta$, however, the partition function is the same up to ${\cal O} (1/N)$ corrections as when ABC are applied. Therefore, regarding orientifold planar equivalence, we argue that in the one-loop approximation ${\cal C}$-breaking in QCD(AS/S,+) resulting from the application of PBC on fermions does not invalidate the large $N$ equivalence with QCD(Adj,-). Similarly, with respect to orbifold planar equivalence, breaking of $Z_2$ interchange symmetry resulting from application of PBC to bifundamental (BF) representation fermions does not invalidate equivalence with QCD(Adj,-) in the one-loop perturbative limit because the partition functions of QCD(BF,-) and QCD(BF,+) are the same. Of particular interest as well is the case of adjoint fermions where for $N_f > 1$ Majorana flavour confinement is obtained for sufficiently small $m \beta$, and deconfinement for sufficiently large $m \beta$. For $N \ge 3$ these two phases are separated by one or more additional phases, some of which can be characterized as partially-confining phases.}


\preprint{}

\begin{document}


\section{\label{introduction}Introduction}
There has been substantial recent interest in properties of QCD-like gauge theories. These field theory models typically involve SU(N) gauge theories with fermions in representations beyond the fundamental, including the adjoint, symmetric, and antisymmetric representations. In addition to the standard antiperiodic boundary conditions applied to fermions there are also good reasons for studying cases in which periodic boundary conditions are applied. For example, $SU(N)$ gauge theories with adjoint fermions, and related theories that preserve the center symmetry and confining properties of the pure gauge theory \cite{Schaden:2005fs,Myers:2007vc,Ogilvie:2007tj,Unsal:2008ch,Myers:2008ey}, have proven to be very useful in understanding mechanisms of confinement \cite{Unsal:2007vu,Unsal:2007jx,Shifman:2008cx,Shifman:2008ja,Ogilvie:2008fz,Shifman:2009tp}. In particular, lattice simulations and analytical results both indicate the existence of a perturbatively accessible confining region, which is analytically connected to the confined phase of the pure $SU(N)$ gauge theory, when periodic boundary conditions are applied to an adjoint potential contribution \cite{Myers:2007vc,Wozar:2008nv}. These theories are also useful in the study of Eguchi-Kawai reduction \cite{Eguchi:1982nm}, in which confined gauge theories should exhibit volume independence in the large $N$ limit \cite{Unsal:2008ch}.

Gauge theories with fermions in higher dimensional representations than the fundamental, such as the two index symmetric, antisymmetric, and adjoint representations are also required in the equivalence between a supersymmetric theory and a non-supersymmetric one, in the large N limit, known as orientifold planar equivalence \cite{Armoni:2003gp,Armoni:2004ub}. These models are also relevant for the construction of conformal \cite{Appelquist:1998xf,Appelquist:1998rb} and near-conformal \cite{Bando:1987we,Cohen:1988sq} QCD-like gauge theories \cite{Catterall:2007yx,Shamir:2008pb,Svetitsky:2008bw,DeGrand:2008dh,DelDebbio:2008tv,Catterall:2008qk,Hietanen:2008vc,Fodor:2008hm,DeGrand:2008kx}.


It has been known for some time that perturbation theory is not valid for calculation of phase transitions from a weakly-coupled phase to a strongly-coupled phase which is clear from high temperature calculations in pure gauge theories and QCD \cite{Gross:1980br}. However, various phases are accessible in QCD-like theories at weak coupling when PBC(+) are applied to fermions. For example, when fermions are in the adjoint representation perturbation theory has revealed an exotic phase structure, and ${\cal C}$-breaking phases are observed for fundamental, symmetric or antisymmetric representation fermions.

It is possible to calculate the phase diagram using high temperature perturbation theory for $SU(N)$ gauge theories with fermions in an arbitrary representation. We refer to these theories as QCD(R), where $R$ is the representation of fermions used. QCD(F,-) corresponds to ordinary QCD when $N = 3$, where the (-) indicates antiperiodic boundary conditions (ABC), in the time dimension, on fermions. When considering periodic boundary conditions (PBC) on fermions we use the symbol (+). The gauge fields will always have the usual periodic boundary conditions applied. Our results reside on the topology $S^1 \times R^3$ where the time dimension is compactified with length $\beta$. When ABC are applied to fermions, this corresponds to taking the theory at finite temperature. \footnote{When PBC are applied to fermions the theory does not correspond to a finite temperature field theory [ABC are required in that case due to the trace over (physical) anticommuting fields in $Z = \Tr \left( e^{- \beta H} \right)$]. However, the formalism is the same so we do not refrain from using the language of thermodynamics, even though it only strictly applies in the case of ABC on fermions.}

In this paper we provide a catalog of some of the phases observable from one-loop perturbation theory in QCD(R,$\pm$) for a single small compact dimension with fermions in the fundamental (F), symmetric (S), anti-symmetric (AS), and adjoint (Adj) representations. The one-loop calculations of $\ln Z_{R,\pm}$ for finite fermion mass $m$, finite chemical potential $\mu$, and constant finite gauge field $A_4$ are included in Appendix \ref{lnZ}, where we provide a detailed derivation which compiles results scattered among various sources: For pure $SU(N)$ gauge theory, $V_{eff}$ has been calculated to one-loop order in \cite{Gross:1980br}. In \cite{Meisinger:2001fi,Kapusta:2006pm} the contribution of fermion mass is included at one-loop for finite $T$ and $\mu$. In \cite{KorthalsAltes:1999cp} the one and two-loop results are calculated for QCD with one flavour of massless fermions and finite $\mu$. In \cite{Unsal:2006pj} the one-loop effective potential was calculated in QCD(AS/A/Adj) for the case $m = 0$ and $\mu = 0$.

The partition function at weak coupling has also been clearly derived at one-loop for $SU(N)$ gauge theories on $S^1 \times S^3$ in the very clear and pedagogical papers \cite{Aharony:2003sx,Aharony:2005bq}.  In \cite{Aharony:2003sx} the authors calculate the large-$N$ deconfining phase transition temperature for the pure Yang-Mills theory (and the Hagedorn transition temperature in ${\cal N} = 4$ SYM) on $S^1 \times S^3$. In \cite{Aharony:2005bq} they show that the Yang-Mills theory deconfinement transition is first-order. Gauge theories on $S^1 \times S^3$ and $S^1\times {\field R}^3$ have different, although slightly overlapping regions of perturbative validity so it is important to study both. To obtain true phase transitions on a finite manifold like $S^1 \times S^3$, it is necessary to take the large $N$ limit. On $S^1 \times {\field R}^3$, this is not necessary, but  perturbative calculations on $S^1 \times {\field R}^3$ with a small number of fermion flavours \footnote{For $N_f = 5$ Majorana flavours the validity of calculations on $S^1 \times {\field R}^3$ extends also to the limit of large $S^1$ \cite{Poppitz:2009uq}.} are only valid in the limit of small $S^1$ where $R_{S^1} \ll \Lambda_{QCD}^{-1}$ (see also the end of \cite{Aharony:2005bq} for a nice discussion of this point).

\section{Conventions and methods}

All of the thermodynamics of QCD(R) can be obtained from the partition function $Z_{QCD(R)}$. For $N_f$ quark flavours in representation $R$, with masses $m_f$, at chemical potential $\mu_f$, and at inverse temperature $\beta = 1/T$, the QCD(R) partition function is given by the (Euclidean space) path integral

\begin{equation}
Z_{QCD(R)} \left( \beta, \mu \right) = \int {\mathrm {\cal D}}A \, e^{-S_{YM} (A)} \int {\mathrm {\cal D}} {\bar \psi} {\mathrm {\cal D}} \psi e^{-\int_0^{\beta} {\mathrm d} \tau \int {\mathrm d}^3 {\bf x} \, {\bar \psi} \left( \Dsl_R - \gamma_0 {\cal M} + M \right) \psi} ,
\end{equation}

\noindent where the gauge field $A_{\mu} (x)= T_R^a A_{\mu}^a (x)$, $\mu = 0, 1, 2, 3$, the $a$ indices are $a = 0, ..., N^2 - 1$, and the $T_R^a$ are the generators of $SU(N)$ in the representation $R$. $\psi$ is a $D(R)$ component vector of fermion fields in the representation $R$ containing $N_f$ anti-commuting 4-spinors $\psi_{f}$. ${\bar \psi}$ contains the corresponding antifermion fields. $M$ is the fermion mass matrix where $(M)_{f f'} = m_f \delta_{f f'}$ and ${\cal M}$ is the fermion chemical potential matrix where $({\cal M})_{f f'} = \mu_f \delta_{f f'}$. $\Dsl_R = \gamma_{\mu} D_{\mu}$ where $D_{\mu}$ is the covariant derivative

\begin{equation}
D_{\mu} = \partial_{\mu} + A_{\mu} .
\end{equation}

\noindent The $A_{\mu}$ are $D(R) \times D(R)$ antihermitian matrices that transform as the representation $R$ of the fermion fields on which they act.

Since our interest is in the phase diagram of this theory we introduce the Polyakov loop order parameter which is defined as the path-ordered exponential of the temporal component of the gauge field,

\begin{equation}
P({\vec x}) = {\cal P} e^{\int_0^{\beta} {\mathrm d}t A_0 (x)} .
\end{equation}

For a constant background field defined by $A_0 \equiv i v / \beta$ the Polyakov loop is

\begin{equation}
P = e^{\beta A_0} = e^{i v} ,
\end{equation}

\noindent where for simplicity we have chosen a gauge in which $A_0$ is diagonal and $v$ is real, diagonal and traceless with elements $(v)_{i j} = v_i \delta_{i j}$. In this case we take $A_0$ (and thus $v$), which transforms as the representation $R$, to be in the form of an $N \times N$ matrix. Then

\begin{equation}
P = {\rm diag} \{ e^{i v_1}, ... , e^{i v_N} \} .
\end{equation}

\subsection{Effective potential}

Details of the derivation of $\ln Z$ on $S^1 \times {\field R}^3$ for arbitrary fermion mass $m$ and chemical potential $\mu$ is included in Appendix \ref{lnZ}. Using this result we obtain the one-loop effective potential in terms of the Polyakov loop,

\begin{equation}
\begin{aligned}
&V_{1-loop} (P, m, \beta, \mu)_{\pm}\\
&= - \frac{1}{\beta V_3} \ln Z (P, m, \beta, \mu)_{\pm}\\
&= \frac{1}{\beta V_3} \left[ - 2 N_f \ln \det \left( - D_R^2 (P) + m^2 \right) + \ln \det \left( - D_{Adj}^{2} (P) \right) \right]\\
&=\frac{m^2 N_f}{\pi^2 \beta^2} \sum_{n=1}^{\infty} \frac{( \pm 1)^n}{n^2} \left[ e^{n \beta \mu} \Tr_R (P^{\dagger n}) + e^{- n \beta \mu} \Tr_R (P^n) \right] K_2 ( n \beta m ) - \frac{2}{\pi^2 \beta^4} \sum_{n=1}^{\infty} \frac{1}{n^4} \Tr_A ( P^n )\\
&= \frac{m^4 N_f}{3 \pi^2} \int_1^{\infty} {\mathrm d}t (t^2 - 1)^{3/2} \left[ g_{R,\pm} \left( \beta, m t, \mu, v \right) + g_{R,\pm}^{\dagger} \left( \beta, m t, - \mu, v \right) \right]\\
&+ \frac{1}{\beta^4} \left[ \frac{1}{24 \pi^2} \sum_{i, j=1}^{N} [ v_i - v_j ]^2 \left( 2 \pi - [ v_i - v_j ] \right)^2 - \frac{\pi^2}{45} \left( N^2 - 1 \right) \right] ,
\end{aligned}
\end{equation}

\noindent where $\Tr_A$ indicates a trace over an object in the adjoint representation, and $g_{R,\pm}$ depends on the group representation of the fermions and the boundary conditions. It is defined as the trace of the matrix

\begin{equation}
\begin{aligned}
g( \beta, m t, \mu, v ) &\equiv \sum_{n=1}^{\infty} e^{- i n v - n \beta (m t - \mu)}\\
&= \frac{e^{- i v} - e^{-\beta (m t - \mu)}}{e^{\beta (m t - \mu)} - 2 \cos v + e^{- \beta (m t - \mu)}}
\end{aligned}
\end{equation}

\noindent in the representation R. The trace in various representations is determined using the Frobenius formula combined with tensor product methods as discussed in Appendix \ref{reps}. We define

\begin{equation}
g_{R,+} \equiv \Tr_R \, g ( \beta, m t, \mu, v )
\end{equation}

\noindent for periodic boundary conditions applied to fermions. The values of $g_{R,\pm}$ for both PBC and ABC applied to fermions can be found in Appendix \ref{lnZ}. $g_{R,-}$ are obtained from $g_{R,+}$ by taking $v \rightarrow v + \pi$.

In what follows we consider the phase diagram of QCD(R) for fermions in the Fundamental (F), symmetric (S), antisymmetric (AS), and adjoint (Adj) representations. We take $\mu =0$ so the result simplifies to

\begin{equation}
\begin{aligned}
&V_{1-loop} (P, m, \beta)_{\pm}\\
&=\frac{2 m^2 N_f}{\pi^2 \beta^2} \sum_{n=1}^{\infty} \frac{(\pm 1)^n}{n^2} {\rm Re} \left[ \Tr_R (P^{n}) \right] K_2 ( n \beta m ) - \frac{2}{\pi^2 \beta^4} \sum_{n=1}^{\infty} \frac{1}{n^4} \Tr_A ( P^n )\\
&= \frac{2 m^4 N_f}{3 \pi^2} \int_1^{\infty} {\mathrm d}t (t^2 - 1)^{3/2} {\rm Re} \left[ g_{R,\pm} \left( \beta, m t, 0, v \right) \right]\\
&+ \frac{1}{\beta^4} \left[ \frac{1}{24 \pi^2} \sum_{i, j=1}^{N} [ v_i - v_j ]^2 \left( 2 \pi - [ v_i - v_j ] \right)^2 - \frac{\pi^2}{45} \left( N^2 - 1 \right) \right]
\end{aligned}
\label{veffmu0}
\end{equation}

\subsection{Chiral condensate}

The chiral (or quark) condensate, $\langle {\bar \psi} \psi \rangle$, is an order parameter for the chiral symmetry of a theory. $\lim_{m \rightarrow 0} \langle {\bar \psi} \psi \rangle \ne 0$ indicates that chiral symmetry is broken. It also serves as an indicator of the order of transitions between phases. From the effective potential it is easy to get the chiral condenstate, which is given by the mass derivative

\begin{equation}
\langle {\bar \psi} \psi \rangle_{1-loop} (m) = - \lim_{V_4 \rightarrow \infty} \frac{1}{V_4 N_f} \frac{\partial}{\partial m} \ln Z(m) = \frac{1}{N_f} \frac{\partial}{\partial m} V_{eff} (P, m) .
\label{chicond}
\end{equation}

\noindent With this definition in QCD(R) for fixed $N$ $\langle {\bar \psi} \psi \rangle$ is independent of $N_f$ if the effective potential is minimized in the same phase (the fermion contribution to $V_{eff}$ is proportional to $N_f$, and this is the only term with mass dependence). In the following sections we measure $\langle {\bar \psi} \psi \rangle$ along with the effective potential.

\section{Results}

To see clearly the effect of fermion mass $m$ on the phase diagram of QCD(R) we perform calculations at $\mu = 0$. In this section we numerically minimize $V_{eff}$ in eq. (\ref{veffmu0}) for $m \beta$ from $0$ to $10$. To ensure accuracy of the results we use a smaller stepsize $\Delta ( m \beta ) = 0.1$ in the region $0 < m \beta \le 3$ since this is where we observe the most phase transitions. The use of a smaller step size also serves as a check that the minimization routine is finding the global minimum. We minimize $V_{eff}$ using an algorithm that searches for the minimum using 10 - 200 random starting points (where more were needed for larger $N$ and complicated phase diagrams) to increase the probability that the global minimum is obtained rather than just a local minimum.

\subsection{Fundamental fermions}

For $SU(N)$ gauge theories with fundamental representation fermions, to which either antiperiodic (-) or periodic (+) boundary conditions have been applied, there are three possible phases which are distinguished in Table \ref{tabF} according to their Polyakov loop eigenvalue angles ${\bf v} = \{ v_1, ..., v_N \}$, as well as $\Tr_F P$.

\begin{table}
\caption{\label{tabF}Table of the possible phases of $SU(N)$ gauge theory with fundamental representation fermions to which either antiperiodic (-) or periodic (+) boundary conditions have been applied, distinguished according to their Polyakov loop eigenvalue angles ${\bf v} \equiv \{ v_1, ..., v_N \}$, as well as $\Tr_F P$. Note $[n,m] \equiv n \mod m$.}
\begin{tabular}{|l|l|l|}
\hline
Phase & ${\bf v}$ & $\Tr_F P$\\
\hline
confined & $\{ \frac{N-1}{N} \pi, \frac{N-3}{N} \pi, ..., - \frac{N-1}{N} \pi \}$ & $0$\\
deconfined & $\{ 0, 0, ..., 0 \}$ & $N$\\
anti-deconfined &  $\{ \pi \pm \frac{\pi}{N} [N,2], \pi \pm \frac{\pi}{N} [N,2], ..., \pi \pm \frac{\pi}{N} [N,2] \}$ & $-N \exp{\left(\pm i \frac{\pi}{N} [N,2] \right)}$\\
\hline
\end{tabular}
\end{table}

The confined and deconfined phases are familiar. The confined phase is that of pure gauge theory. It is located in a region of strong coupling and is thus only accessible non-perturbatively, for example by using lattice simulations. QCD(F) corresponds to pure $SU(N)$ gauge theory when the fundamental fermion mass $m \rightarrow \infty$. In the case of fundamental fermions with non-infinite mass at a value of $\beta = 2 N / g^2$ for which the confined phase is observed in the pure gauge theory, we expect ${\bf v}$ and $\Tr_F P$ to be different from those of the confined phase in Table \ref{tabF} \footnote{This behaviour is suggested in \cite{Myers:2008zm} using 1-loop results for $SU(N)$ gauge theory with both adjoint and fundamental representation fermions (the adjoint fermions are used to make the confined phase accessible perturbatively). The confined phase has ${\bf v} = \{ 0, \phi, - \phi \}$. When the fundamental fermion mass $m_F \rightarrow \infty$, $\phi \rightarrow 2 \pi / 3$ as expected. As $m_F$ is decreased from infinity $\phi$ also decreases.}. This is because fundamental fermions explicitly break the $Z(N)$ center symmetry of the theory. However, ${\bf v}$ and $Tr_F P$ should approach the pure gauge values as $m \rightarrow \infty$.

\subsubsection{Antiperiodic boundary conditions {\protect [ABC(-)]}}

Minimizing the one-loop effective potential of eq. (\ref{veffmu0}) we find that when ABC are applied to fundamental fermions the deconfined phase of Table \ref{tabF} is always favoured. To consider a physical theory, we take $N = 3$ in addition to ABC on fermions, which gives QCD at finite temperature. As expected in the perturbative limit, for all $m \beta$ the effective potential is only minimimized for Polyakov loop angles corresponding to the deconfined phase, such that $\Tr_F P$ is magnetized along the positive real axis:

\begin{equation}
{\bf v} = \{ 0, 0, 0 \}; \hspace{1cm} \Tr_F P = 3 .
\end{equation}

\begin{figure}[t]
  \hfill
  \begin{minipage}[t]{.45\textwidth}
    \begin{center}  
      \includegraphics[width=0.95\textwidth]{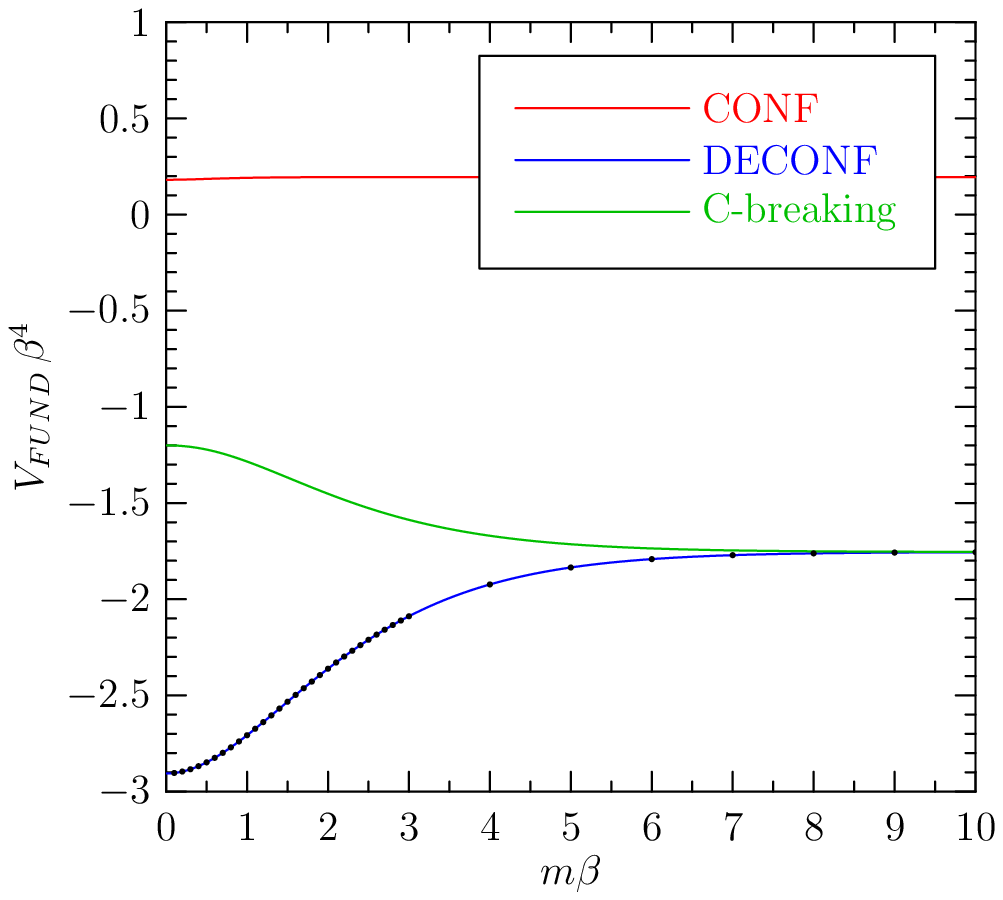}
    \end{center}
  \end{minipage}
  \hfill
  \begin{minipage}[t]{.45\textwidth}
    \begin{center}
\includegraphics[width=0.94\textwidth]{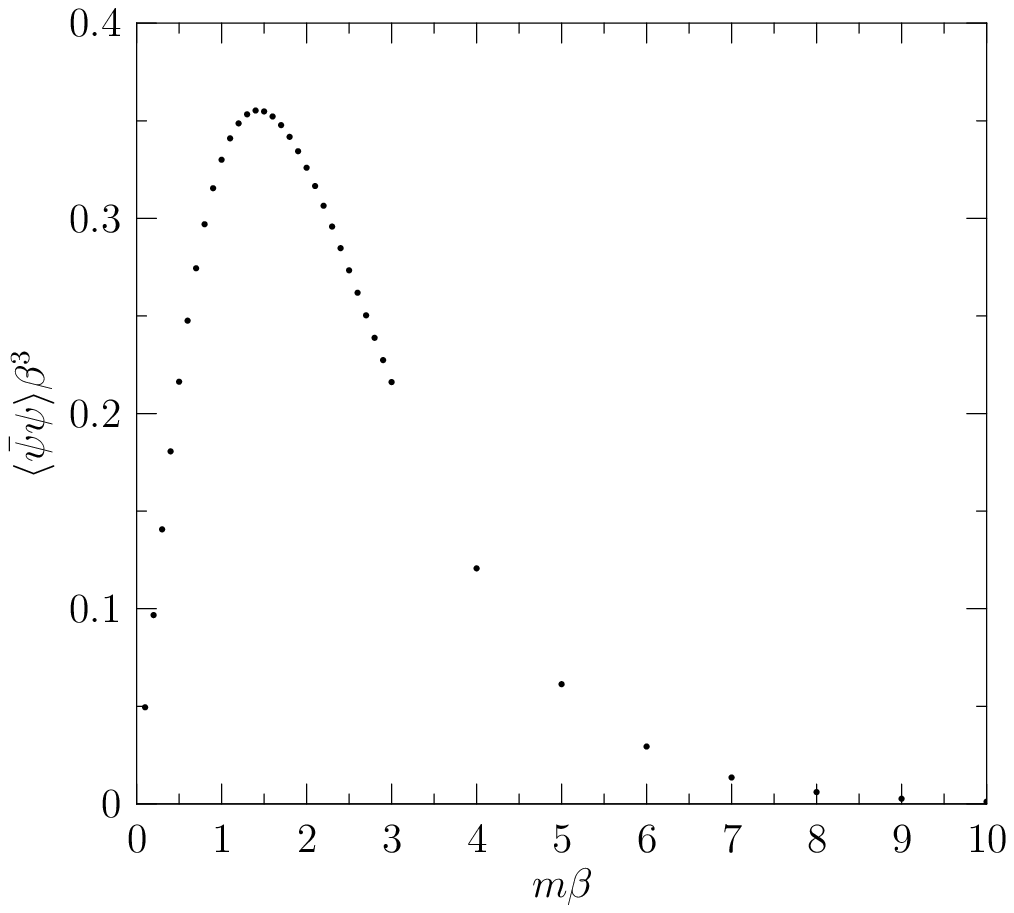}
    \end{center}
  \end{minipage}
  \hfill
  \caption{QCD: (Left) $V_{FUND(-)}$ for $N_f = 1$; (Right) $\langle {{\bar \psi} \psi} \rangle_{FUND(-)}$.}
  \label{qcd_veff}
\end{figure}

The effective potential is plotted in Figure \ref{qcd_veff} (Left). In this type of plot the dots correspond to the minimization of eq. (\ref{veffmu0}) with respect to the Polyakov loop angles $v_i$ for a range of $m \beta$ from $0$ to $10$. The curves correspond to possible phases of QCD distinguished by the values of the Polyakov loop angles as presented in Table \ref{tabF}. Even though only the deconfined phase is accessible via perturbation theory in QCD, a significant feature in Figure \ref{qcd_veff} (Left) is the presence of the inflection point in $V_{eff}$ at $m \beta \approx 1.4$. This implies a large one-loop contribution to the chiral condensate $\langle {{\bar \psi} \psi} \rangle$ as indicated in Figure \ref{qcd_veff} (Right). Including non-perturbative contributions $\langle {\bar \psi} \psi \rangle = 0$ is expected in the high temperature limit of QCD.

\subsubsection{Periodic boundary conditions {\protect [PBC(+)]}}

\begin{figure}[t]
\begin{center}
\includegraphics[width=8cm]{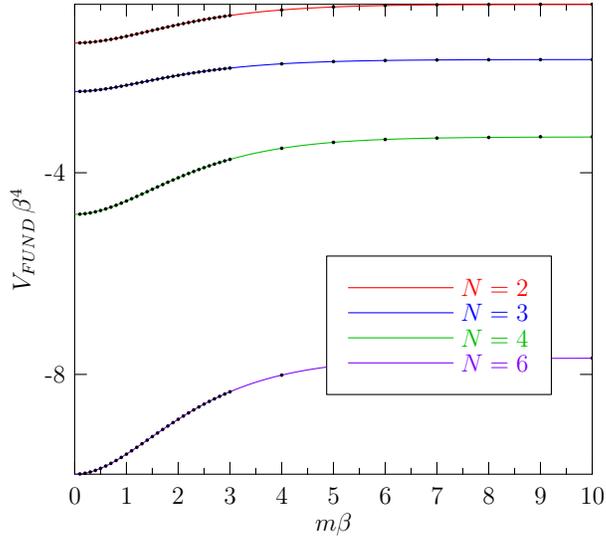}
\end{center}
\caption{$V_{FUND(+)}$ for $N_f = 1$. The dots correspond to minimization of the effective potential of QCD(F,+) with respect to the Polyakov loop angles $v_i$. The curves correspond to evaluation of the effective potential for $v_i = \pi \pm \frac{\pi}{N} [N,2]$, indicating that the anti-deconfined phase of Table {\protect \ref{tabF}} is favoured for all $m \beta$. The curves corresponding to other phases in Table {\protect \ref{tabF}} are not included for clarity, and because they would correspond to larger values of the effective potential.}
\label{v_fund_pbc}
\end{figure}

When PBC are applied to fundamental fermions $V_{FUND(+)}$ always favours the anti-deconfined phase of Table \ref{tabF} as indicated in Figure \ref{v_fund_pbc}. In this case the Polyakov loop angles take the values which are closest to ${\bf v} = \{ \pi, \pi, ..., \pi \}$. As indicated in Table \ref{tabF}, for $N$ even the Polyakov loop angles take those exact values such that $\Tr_F P = - N$, but for $N$ odd they can only come close, taking the values

\begin{equation}
v_i = \pi \pm \frac{\pi}{N} \hspace{7mm} \forall i,
\end{equation}

\noindent which makes $P$ complex. In other words, for $N$ odd the Polyakov loop eigenvalues will take the value of either of the two Nth roots of unity which are closest to $-1$. In this case the ${\cal C}$-symmetry is broken since the Polyakov loop is not invariant under $P \rightarrow P^*$. For the familiar case of $N = 3$ but with PBC on fermions either of the two complex phases corresponding to $v_i = \pm 2 \pi / 3$ are preferred, whereas for ABC the phase with $v_i = 0$ is preferred. All three correspond to deconfined phases of the pure gauge theory. Therefore, when fundamental fermions are added to pure gauge theory in a deconfined phase, preference of a particular vacua depends on the type of boundary conditions applied to the fermions.

\subsection{\label{AS-Sfermions}Antisymmetric (AS) / Symmetric (S) fermions}

For $SU(N)$ gauge theories with fermions in the antisymmetric (AS) or symmetric (S) representation, the possible phases are similar. In the non-perturbative regime, for ABC on fermions, we expect to find a phase which approaches the confined phase in the $m \rightarrow \infty$ limit. This phase is not observed in our one-loop perturbation theory calculations. As expected, for AS/S fermions with ABC, the deconfined phase with ${\bf v} = \{ 0, 0, ..., 0 \}$ always minimizes $V_{eff}$. With $N$ even the deconfined phase can also be defined by ${\bf v} = \{ \pi, \pi, ..., \pi \}$ since the effective potential only includes terms where the eigenvalues are subtracted (boson contribution) or added (fermion contribution). However, with PBC on fermions a phase in which the $v_i$ are all the same, and as close as possible to $\pm \pi/ 2$ is favoured. This is because for tensor representation fermions, $g_{AS(+)}$ and $g_{S(+)}$ of eqs. (\ref{gpbcas}) and (\ref{gpbcs}) depend on $v_i + v_j$, where in the case of fundamental fermions $g_{F(+)}$ of eq. (\ref{gpbcf}) depends on $v_i$, where values closest to $\pi$ were preferred. Since $v_i$ close to $\pm \pi / 2$ is preferred for all $N$ when considering AS/S fermions with PBC, the ${\cal C}$-symmetry is always broken [the exception is $N = 2$ QCD(S) as it is equivalent to QCD(Adj)].

For $N = 4, 8, 12, ...$ the ${\cal C}$-breaking phase has 2 possible vacua corresponding to

\begin{equation}
v_i = \pm \frac{\pi}{2} \hspace{7mm} \forall i.
\label{qcd_as_mult4}
\end{equation}

\noindent Notice that these are exchanged under ${\cal C}$.

For $N$ odd the (double) ${\cal C}$-breaking phase has 4 minima,

\begin{equation}
v_i = \pm \left( \frac{\pi}{2} \pm \frac{\pi}{2 N} \right) \hspace{7mm} \forall i ,
\label{qcd_as_nodd}
\end{equation}

\noindent where $v_i = \pm \left( \frac{\pi}{2} +\frac{\pi}{2 N} \right)$ are exchanged under ${\cal C}$, and $v_i = \pm \left( \frac{\pi}{2} - \frac{\pi}{2 N} \right)$ are exchanged under ${\cal C}$.

\begin{figure}
  \hfill
  \begin{minipage}[t]{.45\textwidth}
    \begin{center}  
      \includegraphics[width=0.95\textwidth]{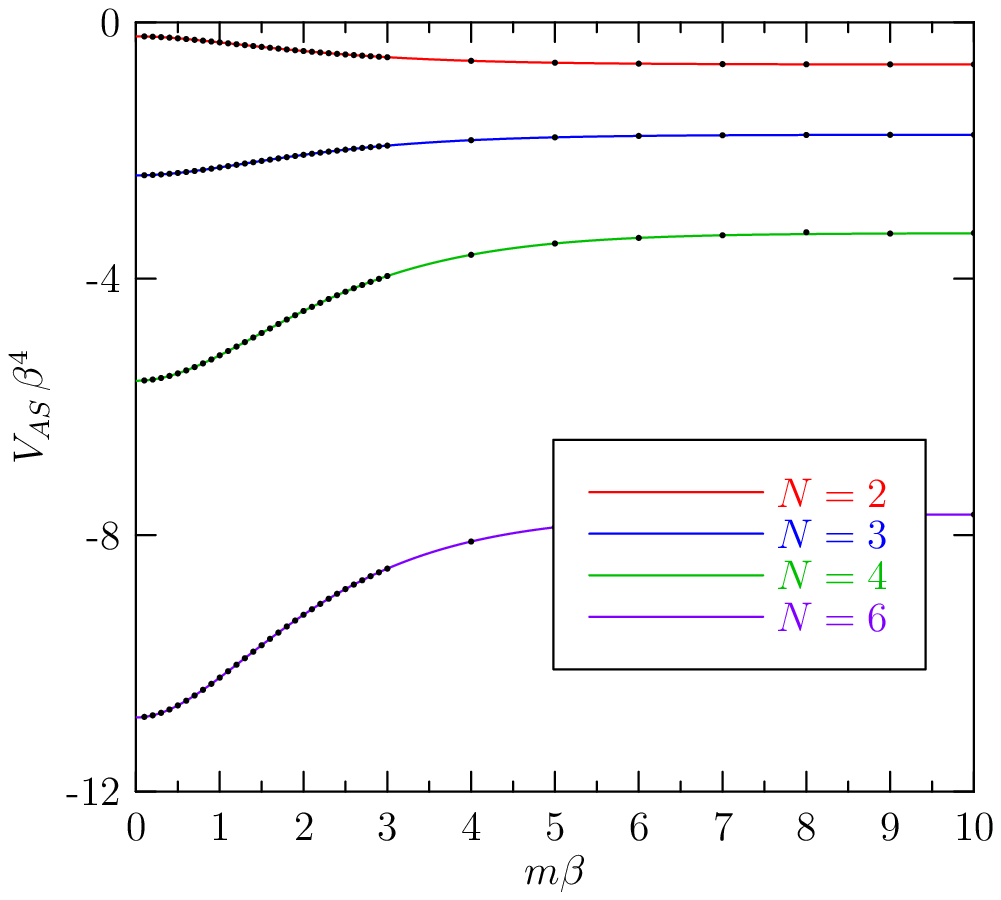}
    \end{center}
  \end{minipage}
  \hfill
  \begin{minipage}[t]{.45\textwidth}
    \begin{center}
\includegraphics[width=0.94\textwidth]{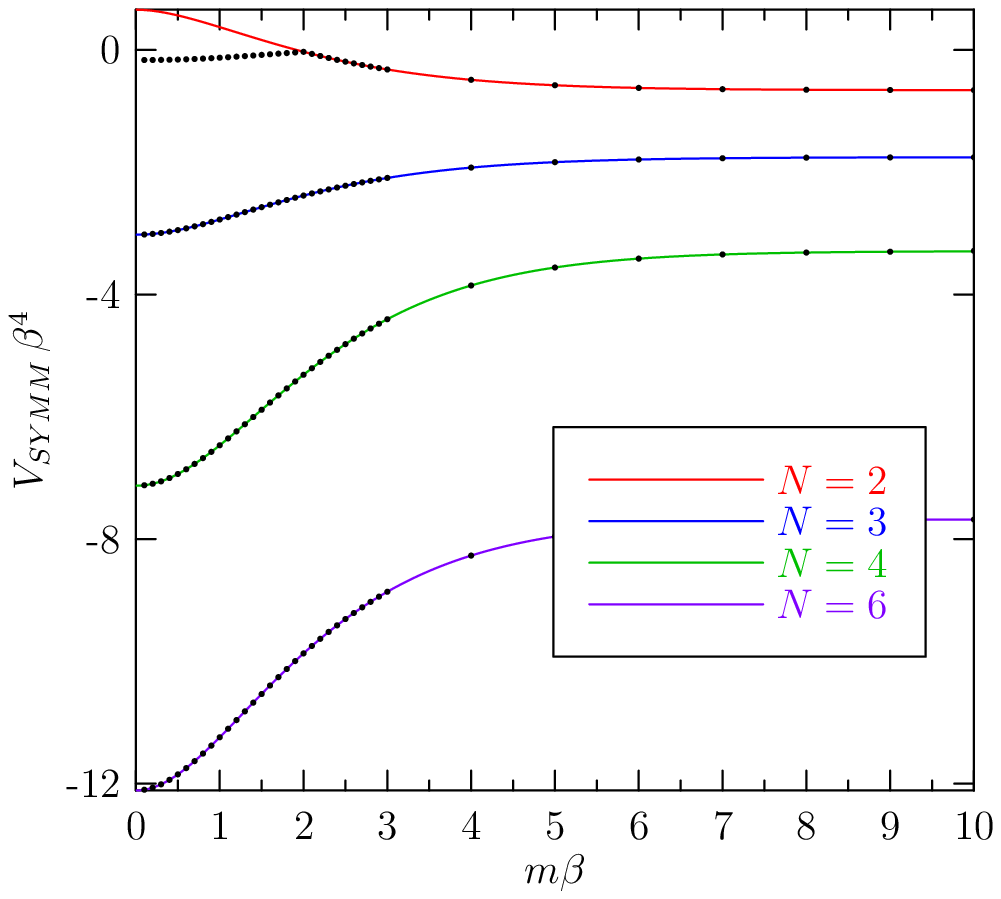}
    \end{center}
  \end{minipage}
  \hfill
  \caption{$N_f = 2$: (Left) $V_{AS(+)}$; (Right) $V_{SYMM(+)}$. The dots correspond to minimization of $V_{eff}$ in eq. ({\protect \ref{veffmu0}}) with respect to the $v_i$. The curves result from evaluation of eq. ({\protect \ref{veffmu0}}) for $v_i = \pm [N,2] \left( \frac{\pi}{2} \pm \frac{\pi}{2 N} \right) \pm \frac{1}{2} [N,4] [(N+1),2] \left( \frac{\pi}{2} \pm \frac{\pi}{N} \right) \pm \frac{1}{2} [(N+2),4][(N+1),2] \left( \frac{\pi}{2} \right)$, as in Table {\protect \ref{tabAS}}. This indicates that in QCD(AS/S,+) for any $N$ the ${\cal C}$-breaking phase is favoured for all $m \beta$, except for $N = 2$ QCD(S), which is equivalent to QCD(Adj) [see also Figure {\protect \ref{pt_adj_nc2_nf2}}].}
  \label{v_as_s_pbc}
\end{figure}

For $N = 2, 6, 10, ...$ the (double) ${\cal C}$-breaking phase has 4 minima,

\begin{equation}
v_i = \pm \left( \frac{\pi}{2} \pm \frac{\pi}{N} \right) \hspace{7mm} \forall i,
\label{qcd_as_nminus2mult4}
\end{equation}

\noindent where $v_i = \pm \left( \frac{\pi}{2} \pm \frac{\pi}{N} \right)$ are exchanged under ${\cal C}$, and $v_i = \pm \left( \frac{\pi}{2} - \frac{\pi}{N} \right)$ are exchanged under ${\cal C}$. The effective potential for various $N$ minimized in the ${\cal C}$-breaking phases is shown in Figure \ref{v_as_s_pbc} (Left) for QCD(AS,+) and Figure \ref{v_as_s_pbc} (Right) for QCD(S,+). The possible phases of $SU(N)$ gauge theory with AS/S representation fermions and either ABC or PBC on fermions are summarized in Table \ref{tabAS}.

\begin{table}
\caption{\label{tabAS}Table of the possible phases of $SU(N)$ gauge theory with antisymmetric (AS) or symmetric (S) representation fermions and either antiperiodic (-) or periodic (+) boundary conditions on fermions. $[n,m] \equiv n \mod m$.}
\begin{tabular}{|l|p{0.8\textwidth}|}
\hline
Phase & ${\bf v}$\\
\hline
confined & $\{ \frac{N-1}{N} \pi, \frac{N-3}{N} \pi, ..., - \frac{N-1}{N} \pi \}$\\
deconfined & $\{ 0, 0, ..., 0 \}$ (for $N$ even this phase can also be defined by $\{ \pi, \pi, ..., \pi \}$)\\
${\cal C}$-breaking &  $v_i = \pm [N,2] \left( \frac{\pi}{2} \pm \frac{\pi}{2 N} \right) \pm \frac{1}{2} [N,4] [(N+1),2] \left( \frac{\pi}{2} \pm \frac{\pi}{N} \right)$ \\
& $\hspace{8mm}\pm \frac{1}{2} [(N+2),4][(N+1),2] \left( \frac{\pi}{2} \right)$\\
\hline
\end{tabular}
\end{table}

\subsection{Adjoint fermions}

QCD(Adj) is unique among the models studied here in several regards. For one, quantities including only adjoint representation traces over the Polyakov loop, $\Tr_A P$, like the partition function $Z_{Adj}$, are clearly invariant under $Z(N)$ transformations [$\Tr_A P = \left| \Tr_F P \right|^2 - 1$ is invariant under $P \rightarrow z P$ for $z \in Z(N)$]. This means that all vacua which are $Z(N)$ rotations of each other are equivalent, as is the case for the pure gauge theory. Since all phases have at least one of the vacua lying on the real axis, we never observe ${\cal C}$-symmetry breaking in QCD(Adj) for either PBC or ABC applied to adjoint fermions. In addition, the phase diagram of QCD(Adj,+) with PBC on fermions becomes quite rich for $N_f \ge 2$ Majorana flavours ($N_f$ Majorana flavours $= 2 N_f$ Dirac flavours). For QCD(Adj) we take $N_f$ to be the number of Majorana flavours since adjoint fermions are their own antiparticles. This means that $N_f \rightarrow N_f / 2$ in eq. (\ref{veffmu0}).

When ABC are applied to adjoint fermions the deconfined phases are always favoured in the minimization of $V_{eff}$. In the deconfined phases the $v_i$ are all the same and correspond to one of the Nth roots of unity, as is the case for the pure $SU(N)$ gauge theory.

\begin{figure}
  \hfill
  \begin{minipage}[t]{.45\textwidth}
    \begin{center}  
      \includegraphics[width=0.95\textwidth]{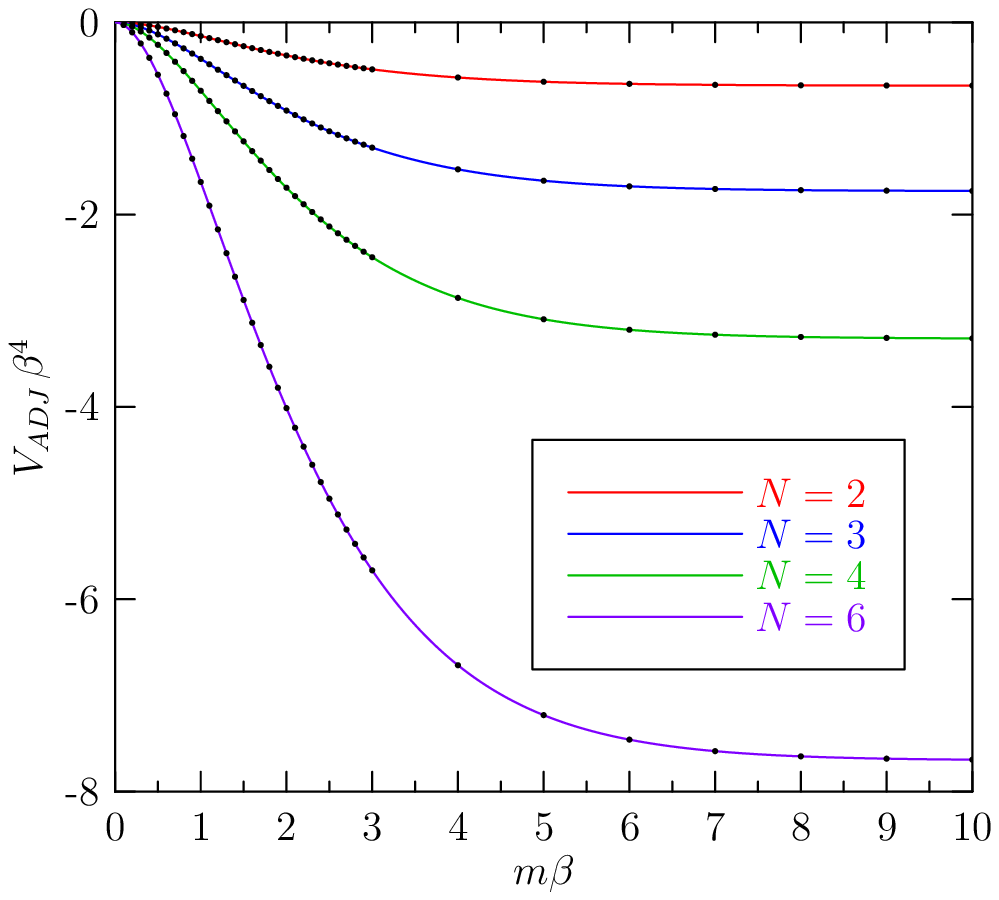}
    \end{center}
  \end{minipage}
  \hfill
  \begin{minipage}[t]{.45\textwidth}
    \begin{center}
\includegraphics[width=0.94\textwidth]{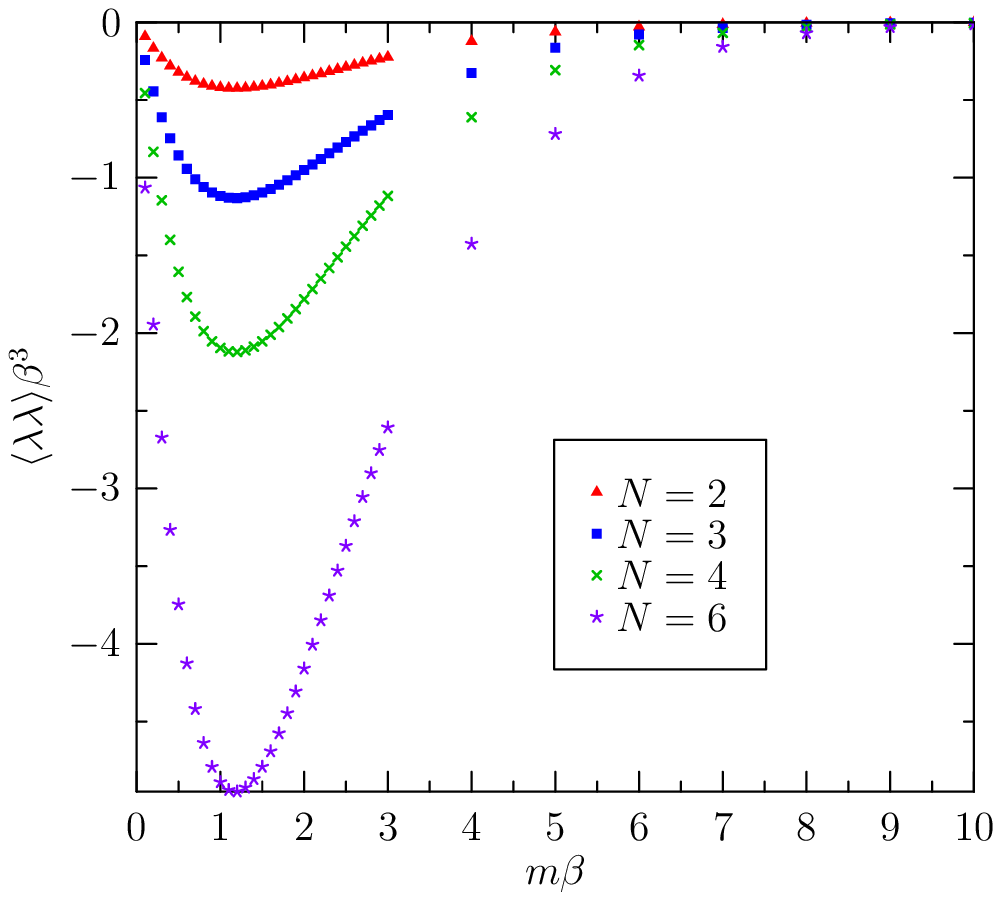}
    \end{center}
  \end{minipage}
  \hfill
\caption{(Left) $V_{ADJ(+)}$ with $N_f = 1$ Majorana flavour. The dots correspond to minimization of $V_{eff}$ in eq. ({\protect \ref{veffmu0}}) with respect to the $v_i$. The curves correspond to the result of evaluating eq. ({\protect \ref{veffmu0}}) for $v_i = 0 \, \forall i$, the values corresponding to the deconfined phase. (Right) $\langle {\lambda \lambda} \rangle_{ADJ(+)}$ for $N_f = 1$ and various values of $N$.}
  \label{adj_veff}
\end{figure}

Changing the fermion boundary conditions to periodic we find that for $N_f = 1$ the deconfined phases are also favoured for all $N$ as shown in Figure \ref{adj_veff} (Left). The corresponding calculation of the chiral condensate, ${\langle \lambda \lambda \rangle}_{1-loop}$ for QCD(Adj,+) is shown in Figure \ref{adj_veff} (Right). However, for $N_f \ge 2$ there is a great variety of observed phases. In particular the confined phase appears to be one of the phases present for all $N$. The observed phases of QCD(Adj,+) for $N_f \ge 2$ and $N$ from $2$ through $9$ are given in Tables \ref{adj_n2} - \ref{adj_n9}. Recall that $Z(N)$ rotations of Polyakov loop eigenvalues included in the tables result in the same value for $V_{eff}$ since it is $Z(N)$-invariant for QCD(Adj).

\begin{table}[p]
\caption{\label{adj_n2}Table of the observed phases $N = 2$ QCD(Adj,+) for $N_f \ge 2$ Majorana flavours.}
\begin{tabular}{|l|l|l|}
\hline
Phase & ${\bf v}$ & $\Tr_F P$\\
\hline
confined & $\{ \frac{\pi}{2}, - \frac{\pi}{2} \}$ & 0 \\
deconfined & $\{ 0, 0 \}$ & 2\\
\hline
\end{tabular}
\end{table}

\begin{table}[p]
\caption{\label{adj_n3}Table of the observed phases $N = 3$ QCD(Adj,+) for $N_f \ge 2$.}
\begin{tabular}{|l|l|l|}
\hline
Phase & ${\bf v}$ & $\Tr_F P$\\
\hline
confined & $\{ \frac{2 \pi}{3}, 0, - \frac{2 \pi}{3} \}$ & 0\\
split & $\{ 0, \pi, \pi \}$ & -1\\
deconfined & $\{ 0, 0, 0 \}$ & 3\\
\hline
\end{tabular}
\end{table}

\begin{table}[p]
\caption{\label{adj_n4}Table of the observed phases $N = 4$ QCD(Adj,+) for $N_f \ge 2$.}
\begin{tabular}{|l|l|l|l|}
\hline
Phase & ${\bf v}$ & $\Tr_F P$ & $\Tr_F P^2$\\
\hline
confined & $\{ \frac{3 \pi}{4}, \frac{\pi}{4}, - \frac{\pi}{4}, - \frac{3 \pi}{4} \}$ & 0 & 0\\
$SU(2)$-confined & $\{ \frac{\pi}{2}, - \frac{\pi}{2}, \frac{\pi}{2}, - \frac{\pi}{2} \}$ & 0 & -4\\
deconfined & $\{ 0, 0, 0, 0 \}$ & 4 & 4\\
\hline
\end{tabular}
\end{table}

\begin{table}[p]
\caption{\label{adj_n5}Table of the observed phases $N = 5$ QCD(Adj,+) for $N_f \ge 2$.}
\begin{tabular}{|l|l|l|}
\hline
Phase & ${\bf v}$ & $\Tr_F P$\\
\hline
confined & $\{ \frac{4 \pi}{5}, \frac{2 \pi}{5}, 0, - \frac{2 \pi}{5}, - \frac{4 \pi}{5} \}$ & $0$\\
attractive & $\{ 0, - \phi, - \phi, \phi, \phi \}$ & $1+ 4 \cos (\phi)$\\
split & $\{ 0, 0, 0, \pi, \pi \}$ & $1$\\
deconfined & $\{ 0, 0, 0, 0, 0 \}$ & $5$\\
\hline
\end{tabular}
\end{table}

\begin{table}[p]
\caption{\label{adj_n6}Table of the observed phases $N = 6$ QCD(Adj,+) for $N_f \ge 2$.}
\begin{tabular}{|l|l|l|l|l|}
\hline
Phase & ${\bf v}$ & $\Tr_F P$ & $\Tr_F P^2$ & $\Tr_F P^3$\\
\hline
confined & $\{ \frac{5 \pi}{6}, \frac{3 \pi}{6}, \frac{\pi}{6}, - \frac{\pi}{6}, - \frac{3 \pi}{6}, - \frac{5 \pi}{6} \}$ & 0 & 0 & 0\\
$SU(3)$-confined & $\{ \frac{2 \pi}{3}, 0, - \frac{2 \pi}{3}, \frac{2 \pi}{3}, 0, - \frac{2 \pi}{3} \}$ & 0 & 0 & 6\\
$SU(2)$-confined& $\{ \frac{\pi}{2}, - \frac{\pi}{2}, \frac{\pi}{2}, - \frac{\pi}{2}, \frac{\pi}{2}, - \frac{\pi}{2} \}$ & 0 & -6 & 0\\
deconfined & $\{ 0, 0, 0, 0, 0, 0 \}$ & 6 & 6 & 6\\
\hline
\end{tabular}
\end{table}

\begin{table}[p]
\caption{\label{adj_n7}Table of the observed phases $N = 7$ QCD(Adj,+) for $N_f \ge 2$.}
\begin{tabular}{|l|l|l|}
\hline
Phase & ${\bf v}$ & $\Tr_F P$\\
\hline
confined & $\{ \frac{6 \pi}{7}, \frac{4 \pi}{7}, \frac{2 \pi}{7}, 0, - \frac{2 \pi}{7}, - \frac{4 \pi}{7}, - \frac{6 \pi}{7} \}$ & 0\\
repulsive & $\{ 0, - \phi, - \phi, \phi, \phi, \pi, \pi \}$ & $-1 + 4 \cos (\phi)$\\
? & $\{ 0, 0, 0, - \phi, - \phi, \phi, \phi \}$ & $3 + 4 \cos (\phi)$\\
attractive & $\{ 0, - \phi, - \phi, - \phi, \phi, \phi, \phi \}$ & $1 + 6 \cos(\phi)$\\
split & $\{ 0, 0, 0, \pi, \pi, \pi, \pi \}$ & $-1$\\
deconfined & $\{ 0, 0, 0, 0, 0, 0, 0 \}$ & $7$\\
\hline
\end{tabular}
\end{table}

\begin{table}[p]
\caption{\label{adj_n8}Table of the observed phases $N = 8$ QCD(Adj,+) for $N_f \ge 2$. Here $c_{(\theta)} \equiv \cos (\theta)$.}
\begin{tabular}{|l|l|l|l|l|}
\hline
Phase & ${\bf v}$ & $\Tr_F P$ & $\Tr_F P^2$ & $\Tr_F P^4$\\
\hline
confined & $\{ \frac{7 \pi}{8}, \frac{5 \pi}{8}, \frac{3 \pi}{8}, \frac{\pi}{8}, - \frac{\pi}{8}, - \frac{3 \pi}{8}, - \frac{5 \pi}{8}, - \frac{7 \pi}{8} \}$ & $0$ & $0$ & $0$\\
$SU(4)$-confined & $\{ \frac{3 \pi}{4}, \frac{\pi}{4}, - \frac{\pi}{4}, - \frac{3 \pi}{4}, \frac{3 \pi}{4}, \frac{\pi}{4}, - \frac{\pi}{4}, - \frac{3 \pi}{4} \}$ & $0$ & $0$ & $-8$\\
attractive & $\{ 0, 0, - \phi, - \phi, - \phi, \phi, \phi, \phi \}$ & $2 + 6 c_{(\phi)}$ & $2 + 6 c_{(2 \phi)}$ & $2 + 6 c_{(4 \phi)}$\\
$SU(2)$-confined & $\{ \frac{\pi}{2}, - \frac{\pi}{2}, \frac{\pi}{2}, - \frac{\pi}{2}, \frac{\pi}{2}, - \frac{\pi}{2}, \frac{\pi}{2}, - \frac{\pi}{2} \}$ & $0$ & $-8$ & $8$\\
deconfined & $\{ 0, 0, 0, 0, 0, 0, 0, 0 \}$ & $8$ & $8$ & $8$\\
\hline
\end{tabular}
\end{table}

\begin{table}[p]
\caption{\label{adj_n9}Table of the observed phases $N = 9$ QCD(Adj,+) for $N_f \ge 2$. Here $c_{(\theta)} \equiv \cos (\theta)$.}
\begin{tabular}{|l|l|l|l|}
\hline
Phase & ${\bf v}$ & $\Tr_F P$ & $\Tr_F P^3$\\
\hline
confined & $\{ \frac{8 \pi}{9}, \frac{6 \pi}{9}, ... , 0, ... , - \frac{6 \pi}{9}, - \frac{8 \pi}{9} \}$ & $0$ & $0$\\
mixed & $\{ 0, - \phi, - \phi, \phi, \phi, - \chi, - \chi, \chi, \chi \}$ & $1 + 4 [ c_{(\phi)} + c_{(\chi)} ]$ & $1 + 4 [ c_{(3 \phi)} + c_{(3 \chi)} ]$\\
$SU(3)$-confined & $\{ \frac{2 \pi}{3}, 0, \frac{4 \pi}{3}, \frac{2 \pi}{3}, 0, \frac{4 \pi}{3}, \frac{2 \pi}{3}, 0, \frac{4 \pi}{3} \}$ & $0$ & $9$\\
attractive & $\{ 0, - \phi, - \phi, - \phi, - \phi, \phi, \phi, \phi, \phi \}$ & $1 + 8 \cos(\phi)$ & $1 + 8 \cos(3 \phi)$\\
split & $\{ 0, 0, 0, 0, 0, \pi, \pi, \pi, \pi \}$ & $1$ & $9$\\
deconfined & $\{ 0, 0, 0, 0, 0, 0, 0, 0 \}$ & $9$ & $9$\\
\hline
\end{tabular}
\end{table}

\clearpage

\begin{figure}[p]
  \hfill
    \begin{minipage}[t]{.45\textwidth}
    \begin{center}
\includegraphics[width=0.9\textwidth]{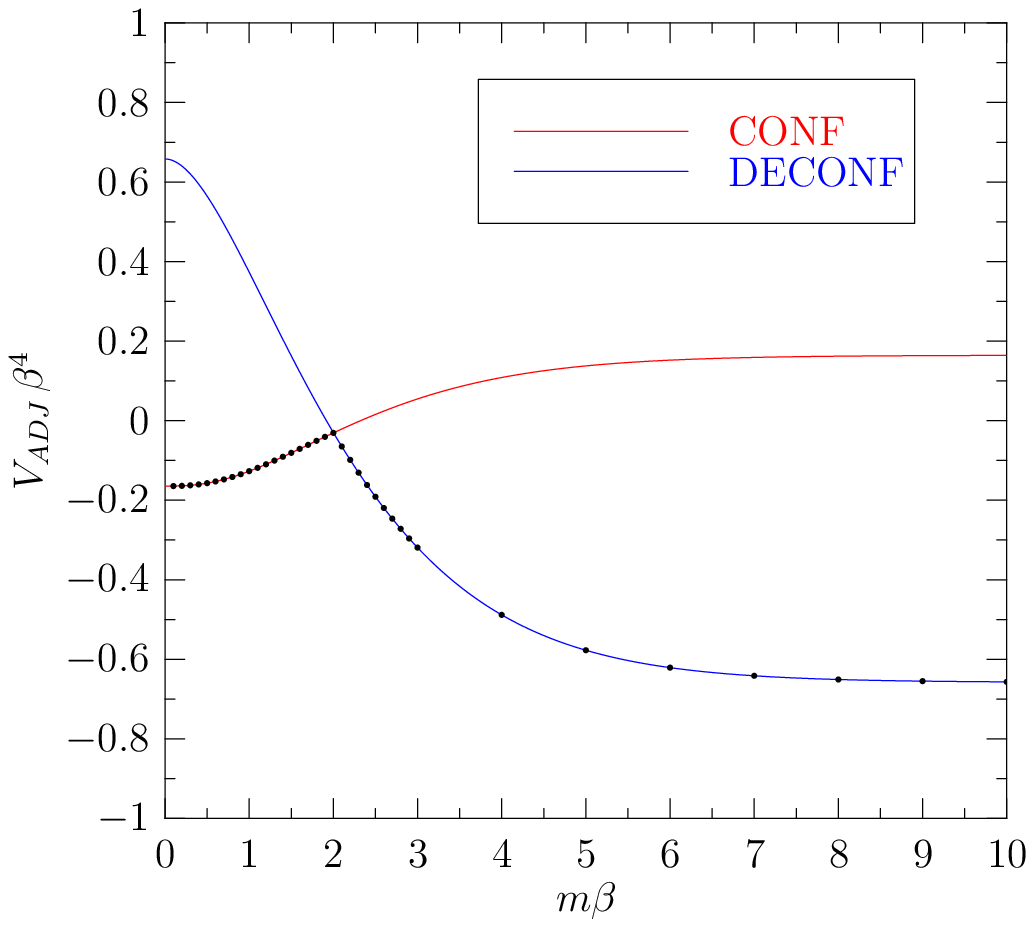}
       \caption{$N = 2$: $V_{ADJ(+)}$ for $N_f = 2$ Majorana or $V_{SYMM(+)}$ for $N_f = 1$ Dirac flavour}
       \label{pt_adj_nc2_nf2}
    \end{center}
  \end{minipage}
  \hfill
  \begin{minipage}[t]{.45\textwidth}
    \begin{center}
\includegraphics[width=0.9\textwidth]{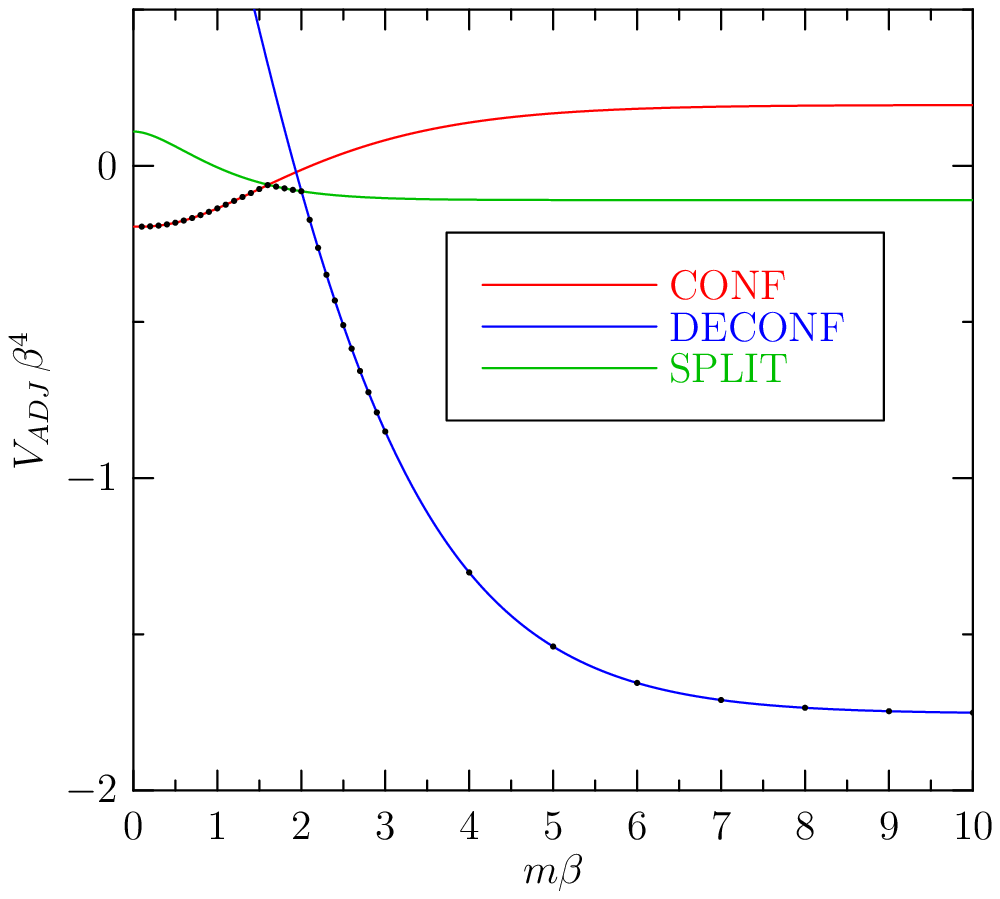}
\caption{$V_{ADJ(+)}$ for $N = 3$ and $N_f = 2$}
\label{pt_adj_nc3_nf2}
    \end{center}
  \end{minipage}
  \hfill
\end{figure}

\begin{figure}[p]
  \hfill
    \begin{minipage}[t]{.45\textwidth}
    \begin{center}
\includegraphics[width=0.9\textwidth]{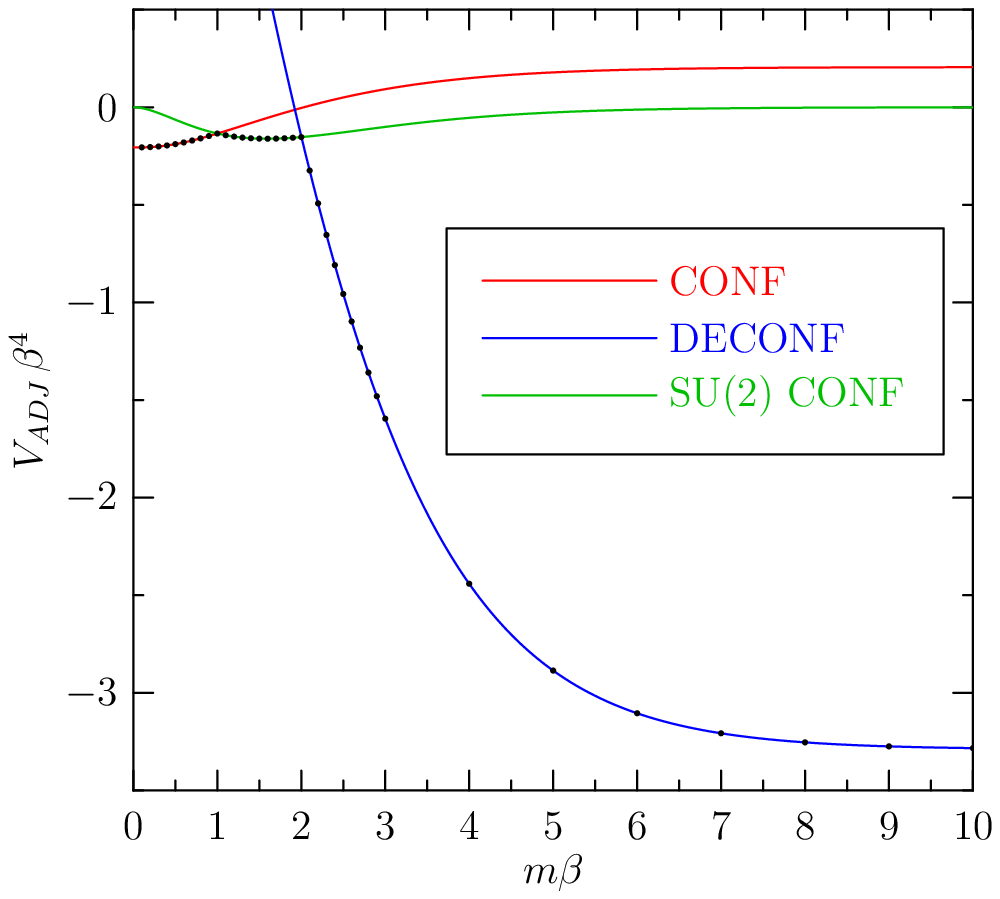}
       \caption{$V_{ADJ(+)}$ for $N = 4$ and $N_f = 2$}
       \label{pt_adj_nc4_nf2}
    \end{center}
  \end{minipage}
  \hfill
  \begin{minipage}[t]{.45\textwidth}
    \begin{center}
\includegraphics[width=0.9\textwidth]{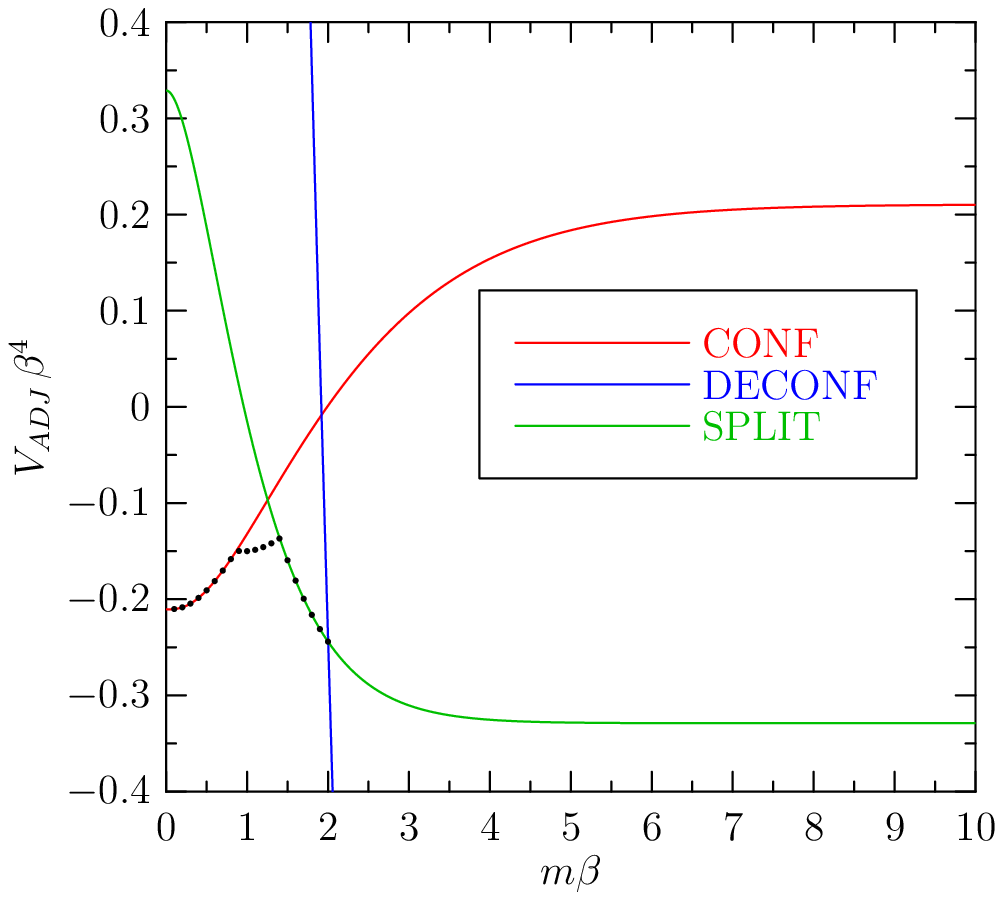}
\caption{$V_{ADJ(+)}$ for $N = 5$ and $N_f = 2$}
\label{pt_adj_nc5_nf2}
    \end{center}
  \end{minipage}
  \hfill
\end{figure}

\begin{figure}[p]
  \hfill
    \begin{minipage}[t]{.45\textwidth}
    \begin{center}
\includegraphics[width=0.9\textwidth]{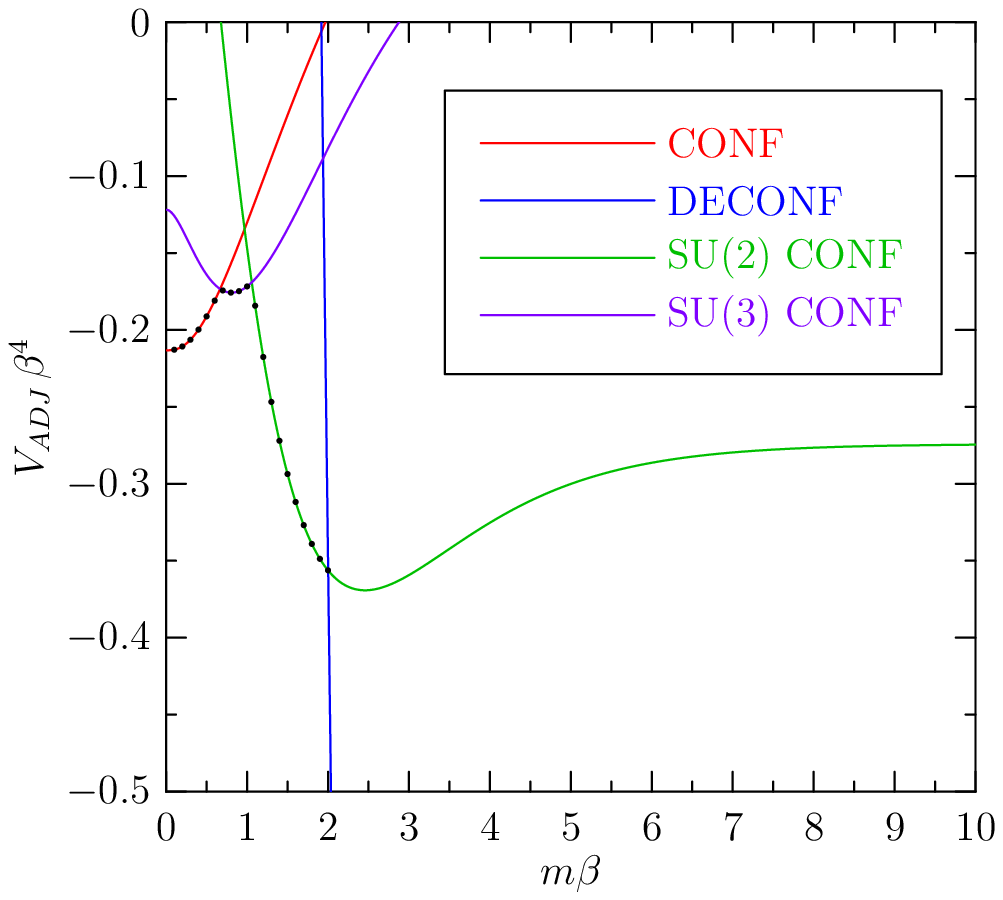}
       \caption{$V_{ADJ(+)}$ for $N = 6$ and $N_f = 2$}
       \label{pt_adj_nc6_nf2}
    \end{center}
  \end{minipage}
  \hfill
  \begin{minipage}[t]{.45\textwidth}
    \begin{center}
\includegraphics[width=0.9\textwidth]{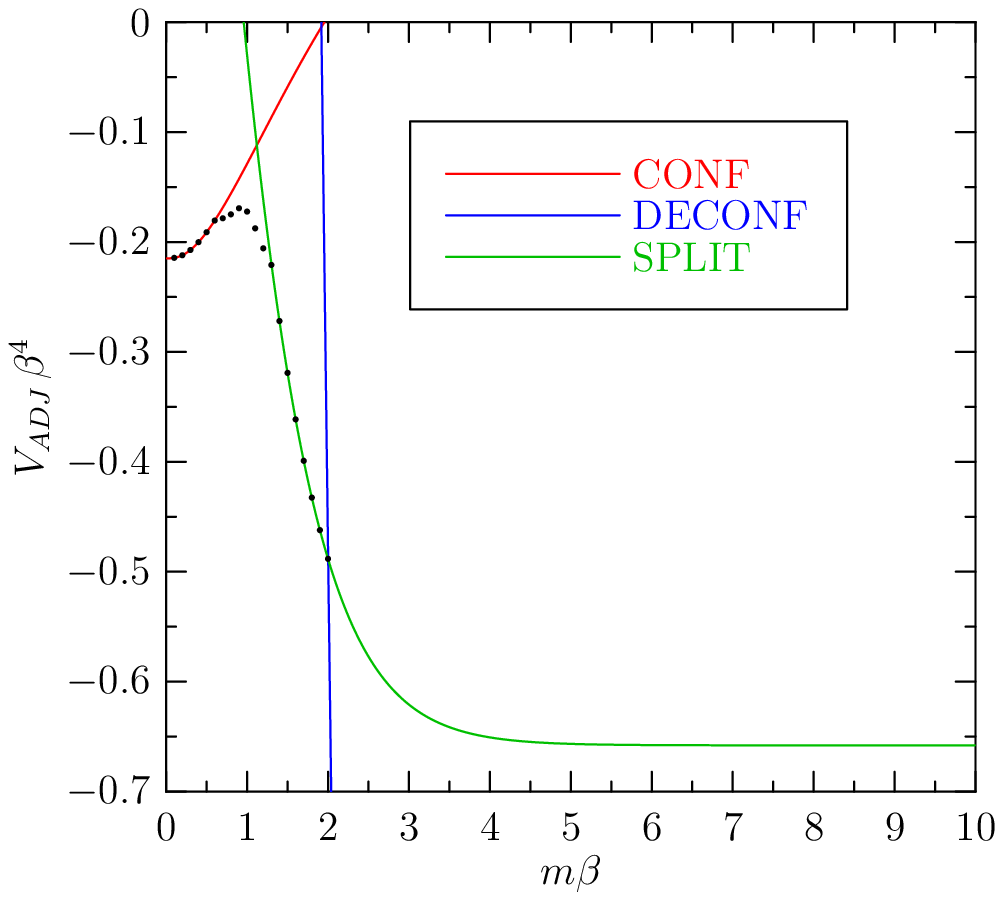}
\caption{$V_{ADJ(+)}$ for $N = 7$ and $N_f = 2$}
\label{pt_adj_nc7_nf2}
    \end{center}
  \end{minipage}
  \hfill
\end{figure}

\begin{figure}[p]
  \hfill
    \begin{minipage}[t]{.45\textwidth}
    \begin{center}
\includegraphics[width=0.9\textwidth]{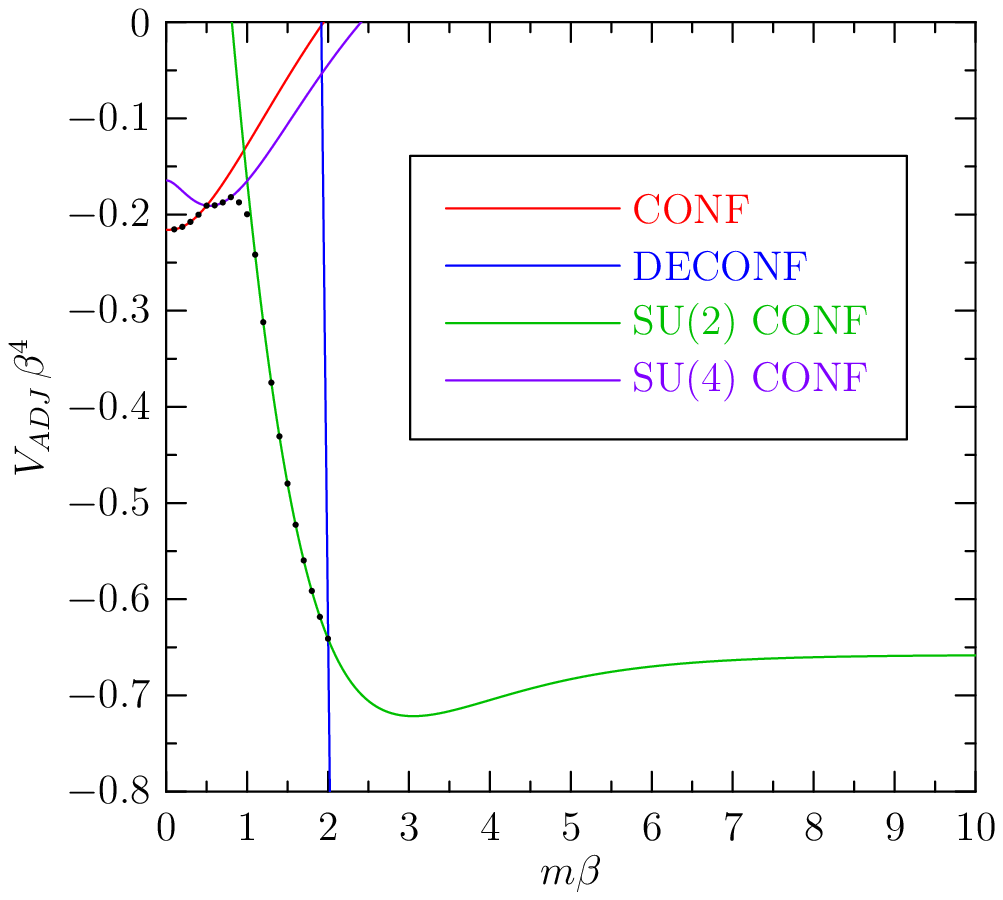}
       \caption{$V_{ADJ(+)}$ for $N = 8$ and $N_f = 2$}
       \label{pt_adj_nc8_nf2}
    \end{center}
  \end{minipage}
  \hfill
  \begin{minipage}[t]{.45\textwidth}
    \begin{center}
\includegraphics[width=0.9\textwidth]{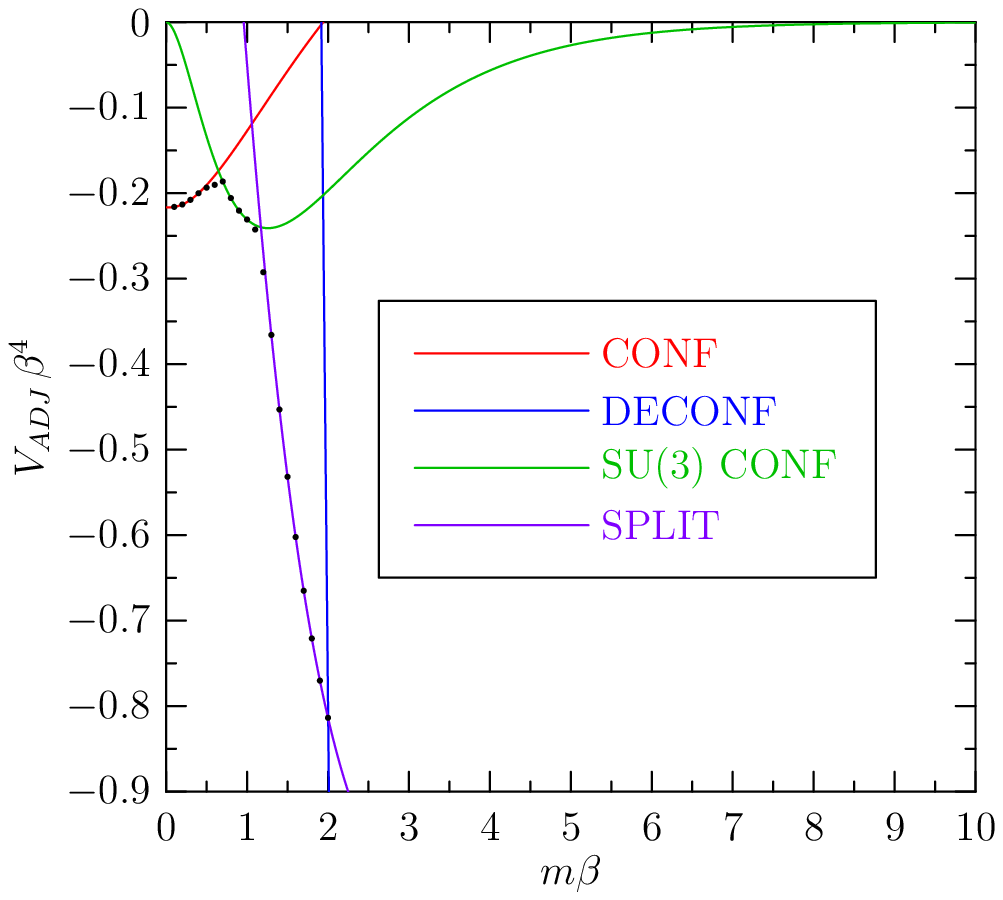}
\caption{$V_{ADJ(+)}$ for $N = 9$ and $N_f = 2$}
\label{pt_adj_nc9_nf2}
    \end{center}
  \end{minipage}
  \hfill
\end{figure}

The results from minimization of the effective potential for QCD(Adj,+) with respect to the Polyakov loop angles $v_i$ for $N_f = 2$ and $N$ from $2$ - $9$ are shown in Figures \ref{pt_adj_nc2_nf2} - \ref{pt_adj_nc9_nf2}. These phase diagrams are quite rich and appear to increase in complexity as $N$ increases. The confined phase is defined by

\begin{equation}
{\bf v} = \{ \frac{N-1}{N} \pi, \frac{N-3}{N} \pi, ..., - \frac{N-1}{N} \pi \}; \hspace{1cm} \Tr_F P = 0.
\end{equation}

\noindent The deconfined phases are defined by

\begin{equation}
{\bf v} = \{ 0, 0, ..., 0 \}; \hspace{1cm} \Tr_F P = N,
\end{equation}

\noindent and $Z(N)$ rotations.

For $N$ not prime there are additional phases which correspond to confinement of a subgroup $L$ of $SU(N)$, which are called $SU(L)$-confined \footnote{Partially confined phases are also discussed in \cite{Myers:2007vc,Ogilvie:2007tj}}. In these phases quarks are still confined, however groups of $L$ quarks are not. The Polyakov loop angles correspond to $N / L$ copies of the $L$ angles of the $SU(L)$ confined phase. For example, in $SU(6)$ we observe $SU(2)$ and $SU(3)$ confined phases. The $SU(2)$ confined phase corresponds to

\begin{equation}
{\bf v} = \{ \frac{\pi}{2}, - \frac{\pi}{2}, \frac{\pi}{2}, - \frac{\pi}{2}, \frac{\pi}{2}, - \frac{\pi}{2} \} ,
\end{equation}

\noindent and the $SU(3)$ confined phase has

\begin{equation}
{\bf v} = \{ \frac{2 \pi}{3}, 0, - \frac{2 \pi}{3}, \frac{2 \pi}{3}, 0, - \frac{2 \pi}{3} \} ,
\end{equation}

\noindent as indicated in Table \ref{adj_n6}. These phases can be distinguished by taking the trace of powers of the Polyakov loop. In the $SU(2)$ confined phase $\Tr_F P = \Tr_F P^3 = 0$, however $\Tr_F P^2 \ne 0$. In the $SU(3)$ confined phase $\Tr_F P = \Tr_F P^2 = 0$, but $\Tr_F P^3 \ne 0$. The same pattern holds for arbitrary $L$.

In addition, for $N = 5, 7, 8, 9$ we find an attractive phase, called such because $\Tr_F P$ is not constant with respect to small changes in $m \beta$, rather $\left| \Tr_F P \right|$ slowly decreases as $m \beta$ increases, indicating that the Polyakov loop eigenvalues are attracted together. In the repulsive phase of $SU(7)$ the Polyakov loop eigenvalues are repelled apart with increasing $m \beta$. In $SU(9)$ there is a mixed phase where some of the eigenvalues attract, while others repel.

For $N$ odd we also find {\it split} phases which favour a small magnitude for $\left| \Tr_F P \right|$. These have $(N \pm 1) / 2$ angles that are $\pi$ and $(N \mp 1) / 2$ angles which are $0$, where the top sign corresponds to $N = 3, 7, 11, ...$ and the bottom sign corresponds to $N = 5, 9, 13, ...$, such that $\det P = 1$ as required. The $Z(N)$ symmetry is completely broken in these phases, albeit in a special way. For example, in $SU(5)$ this phase is given by ${\bf v} = \{ 0, 0, 0, \pi, \pi \}$ which is an $SU(3) \times SU(2)$ [mod $Z(5)$] phase.

In \cite{Myers:2007vc} we observed the $N = 3$ split phase in lattice simulations of a related theory and referred to the phase as "skewed" in that ${\rm Proj}_{Z(N)} [ \Tr_F P] < 0$ as opposed to the deconfined phase where ${\rm Proj}_{Z(N)} [ \Tr_F P] > 0$, causing the vacua of the skewed phases to lie at angles midway between those of the deconfined phases. However, for $N = 5, 9, 13, ...$ the vacua of the split phases have angles corresponding to the $N$ roots of unity, as do the deconfined phases. As will become important shortly, the split phase of odd $N$ and the $SU(2)$-confined phase of even $N$ are related in that $V_{Adj(+)}$ is the same in these phases, in terms of $N$.

It is important to note that additional phases are also possible. Our step size in the region $0.1 \le m \beta \le 3.0$ is $\Delta ( m \beta ) = 0.1$, and for $3 \le m \beta \le 10$ the step size is $\Delta ( m \beta ) = 1$. So there is room for additional phases in narrow regions of $m \beta$ that extend less than $0.1$ for $m \beta \le 3.0$, and less than 1.0 for $3.0 \le m \beta \le 10$.

\subsubsection{Large N limit}

\begin{figure}[t]
\begin{center}
\includegraphics[width=8cm]{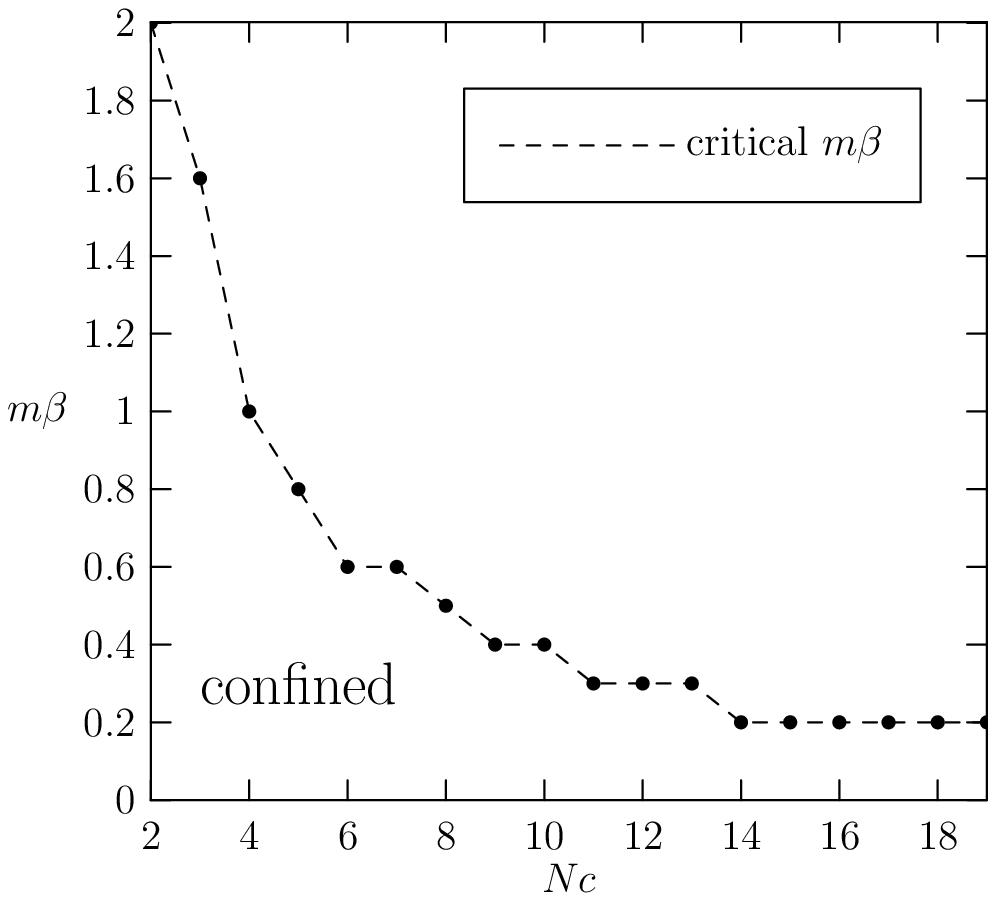}
\end{center}
\caption{Accessibility of the confined phase in QCD(Adj,+) for $N_f = 2$ and $N$ from $2$ - $19$. The curve for $\left( m \beta \right)_{crit}$ is a lower bound on the value of transition out of the confined phase. However, for each $N$, the point at $m \beta = \left( m \beta \right)_{crit} + 0.1$ does not belong to the confined phase.}
\label{conf}
\end{figure}

From the phase diagrams of QCD(Adj,+) in Figures \ref{pt_adj_nc2_nf2} - \ref{pt_adj_nc9_nf2} it is clear that the confined phase extends for a smaller range of $m \beta$ as $N$ increases. To determine if the confined phase persists into the limit of very large $N$ we minimized $V_{eff}$  in the region of small $m \beta$ for values of $N$ up to $19$. The value of $m \beta$ at which there is a transition out of the confined phase was measured. The results are plotted in Figure \ref{conf}. The curve represents a lower bound: for each value of $N$ on the curve the value of $m \beta$ 0.1 above it does not belong to the confined phase. The confined phase becomes quickly less accessible as $N$ increases from $2$ to $10$, but appears to level off in the large $N$ limit providing a narrow region, near $m \beta \sim 0$, in which it is the preferred phase.

Another noteworthy feature of the phase diagrams in Figures \ref{pt_adj_nc2_nf2} - \ref{pt_adj_nc9_nf2} is that the transition to the deconfined phase always occurs at $m \beta = 2.00199$, independent of $N$. This is explained in the next subsection.

\subsubsection{Variations in $N_f$}

\begin{figure}[t]
  \hfill
  \begin{minipage}[t]{.45\textwidth}
    \begin{center}  
      \includegraphics[width=0.95\textwidth]{pt_adj_nc6_nf2.eps}
    \end{center}
  \end{minipage}
  \hfill
  \begin{minipage}[t]{.45\textwidth}
    \begin{center}
\includegraphics[width=0.94\textwidth]{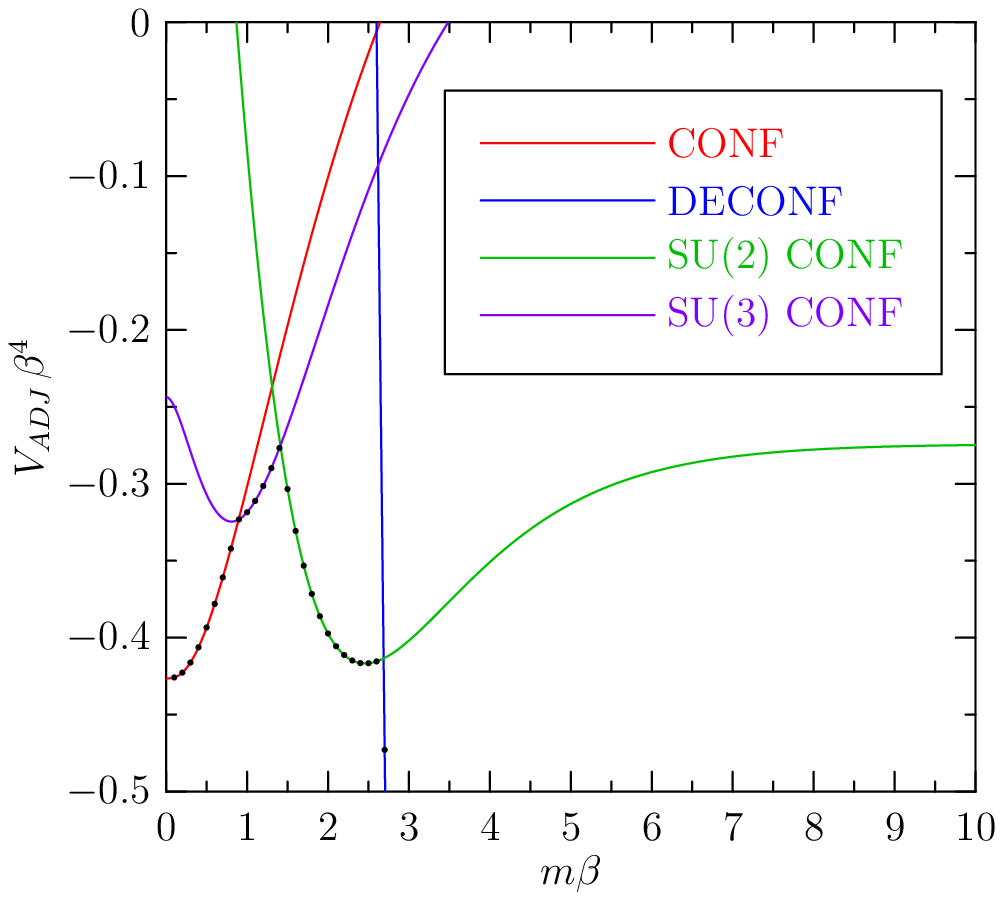}
    \end{center}
  \end{minipage}
  \hfill
  \caption{$V_{ADJ(+)}$ for $N = 6$: (Left) $N_f = 2$; (Right) $N_f = 3$. These results suggest that increasing $N_f$ causes the confined phase to be preferred up to a larger $(m \beta)_{crit}$.}
  \label{adj_change_nf}
\end{figure}

We also calculated the phase diagrams in QCD(Adj,+) for $N_f = 3$ flavours of adjoint fermions for $N = 3, 4, 5, 6$. The general trend is that the value of $m \beta$ below which the confined phase is preferred, given by $(m \beta)_{crit}$, increases with $N_f$. This is illustrated in Figure \ref{adj_change_nf}.

Overall, increasing $N_f$ can offset the decrease in $(m \beta)_{crit}$ with $N$, but there is a natural limit in the value of $N_f$ in that to preserve asymptotic freedom in QCD(Adj) it is necessary to keep $N_f \le 5$ Majorana flavours.

\begin{figure}[t]
\begin{center}
\includegraphics[width=8cm]{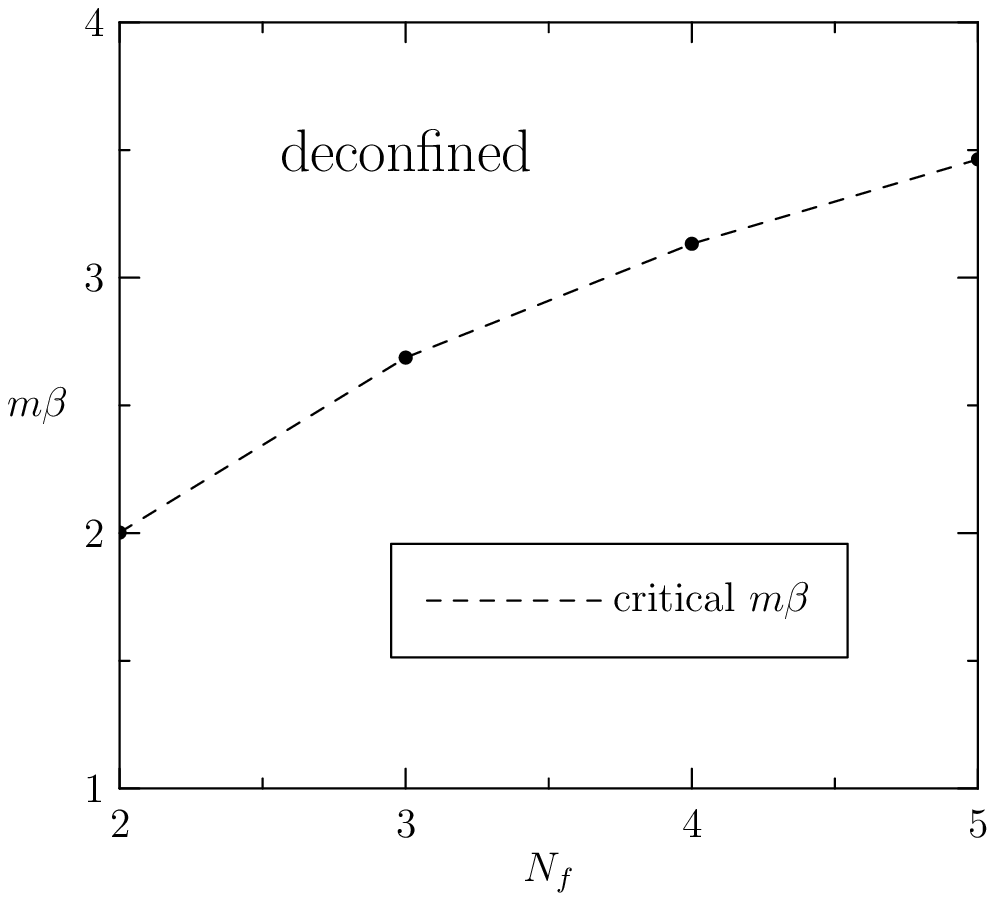}
\end{center}
\caption{Location of the transition to the deconfined phase phase in QCD(Adj,+) for $N_f$ from $2$ through $5$, for all $N$.}
\label{deconf_trans}
\end{figure}

Also, as $N_f$ increases the transition to the deconfined phase, which is still independent of $N$, also increases, as shown in Figure \ref{deconf_trans}. The value of $m \beta$ at which this transition is realized is given by the solution of

\begin{equation}
\sum_{n=1}^{\infty} \frac{1}{( 2 n - 1 )^2} \left[ N_f K_2 \left[ (2 n - 1) m \beta \right] - \frac{2}{\left[ (2 n - 1) m \beta \right]^2} \right] = 0 .
\label{dec_trans_loc}
\end{equation}

\noindent As $m \beta$ is increased the transition to the deconfined phase always occurs from an $SU(2)$-confined phase (for $N$ even) or a split phase (for $N$ odd). The formulas for $V_{Adj(+)}$ are identical in the split and $SU(2)$-confined phases since in both cases $\Tr_A (P^n) = 0$ for $n$ odd and $\Tr_A (P^n) = N^2 - 1$ for $n$ even (see Tables \ref{adj_n2} - \ref{adj_n9}). Since $\Tr_A (P^n) = N^2 - 1$ for all $n$ in the deconfined phase the only terms that differ when comparing the split, or $SU(2)$-confined phase, and the deconfined phase, occur for $n$ odd. Setting the formulas for $V_{Adj(+)}$ in the split (or $SU(2)$-confined) phase and the deconfined phase equal then gives the equality in eq. (\ref{dec_trans_loc}).

\subsubsection{Chiral condensate}

\begin{figure}[p]
  \hfill
    \begin{minipage}[t]{.45\textwidth}
    \begin{center}
\includegraphics[width=0.9\textwidth]{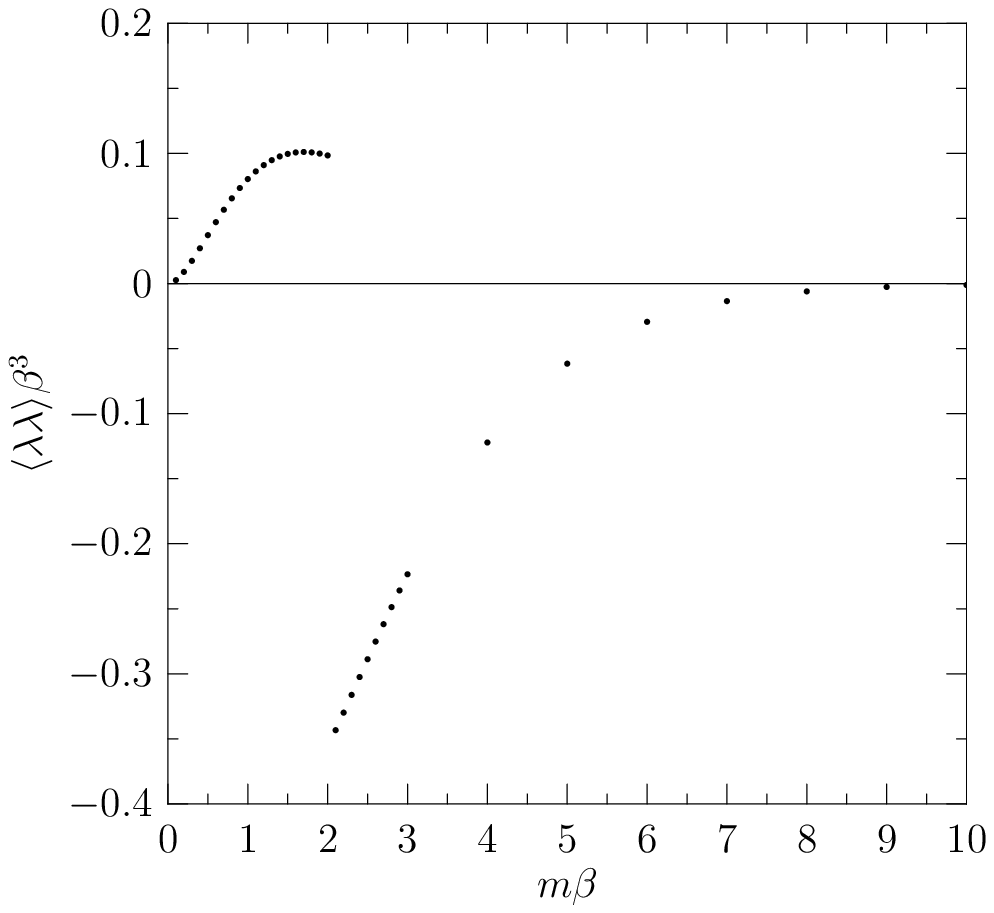}
       \caption{$\langle \lambda \lambda \rangle_{ADJ(+)}$ for $N = 2$, $N_f = 2$}
       \label{chi_adj_nc2_nf2}
    \end{center}
  \end{minipage}
  \hfill
  \begin{minipage}[t]{.45\textwidth}
    \begin{center}
\includegraphics[width=0.9\textwidth]{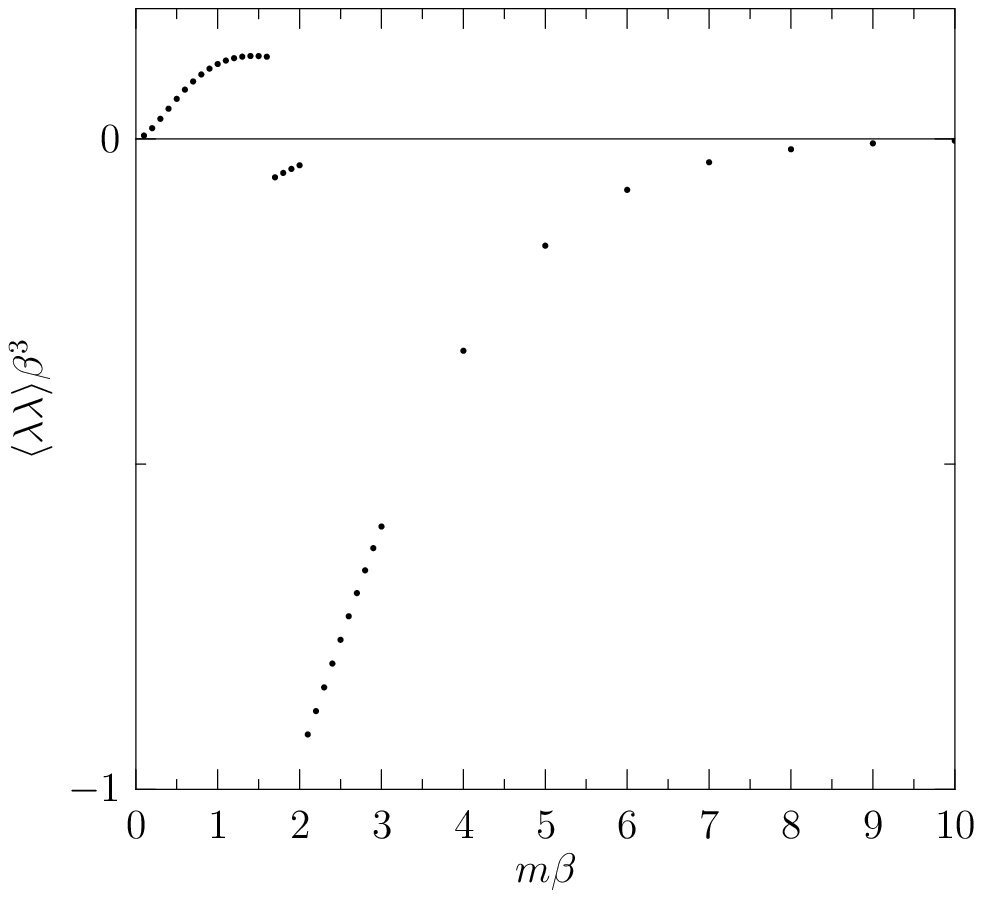}
\caption{$\langle \lambda \lambda \rangle_{ADJ(+)}$ for $N = 3$, $N_f = 2$}
\label{chi_adj_nc3_nf2}
    \end{center}
  \end{minipage}
  \hfill
\end{figure}

\begin{figure}[p]
  \hfill
    \begin{minipage}[t]{.45\textwidth}
    \begin{center}
\includegraphics[width=0.9\textwidth]{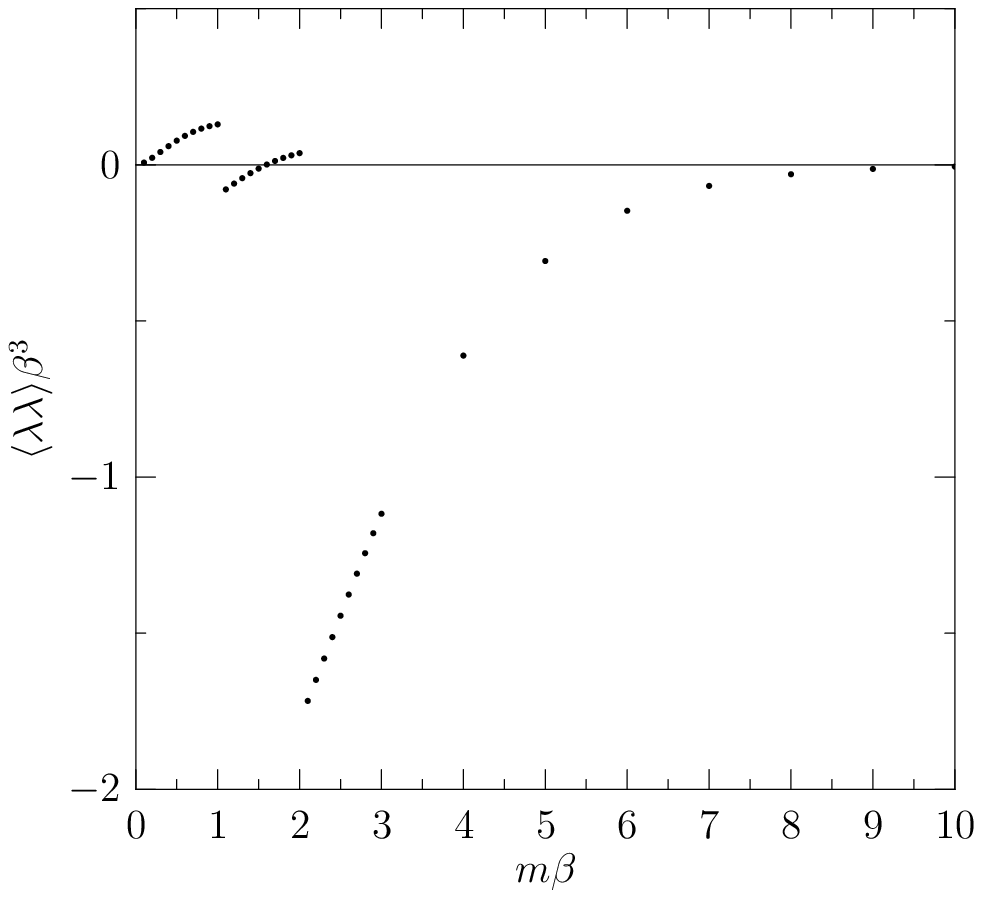}
       \caption{$\langle \lambda \lambda \rangle_{ADJ(+)}$ for $N = 4$, $N_f = 2$}
       \label{chi_adj_nc4_nf2}
    \end{center}
  \end{minipage}
  \hfill
  \begin{minipage}[t]{.45\textwidth}
    \begin{center}
\includegraphics[width=0.9\textwidth]{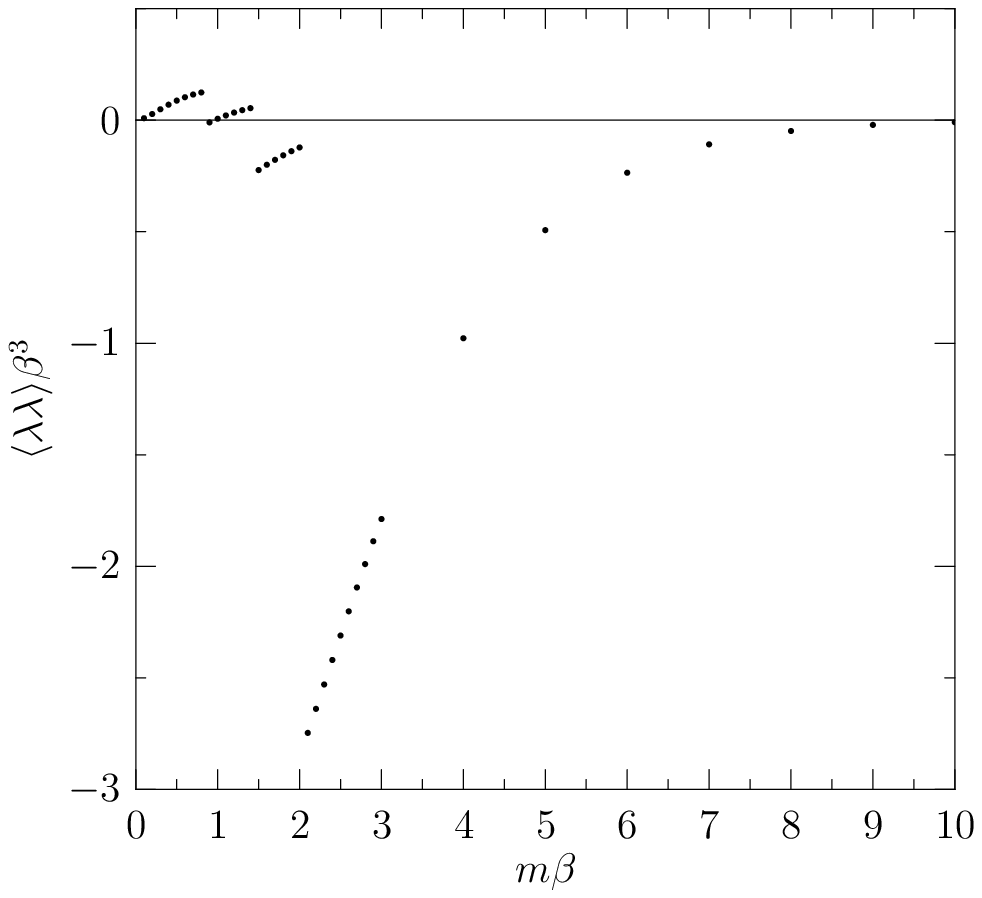}
\caption{$\langle \lambda \lambda \rangle_{ADJ(+)}$ for $N = 5$, $N_f = 2$}
\label{chi_adj_nc5_nf2}
    \end{center}
  \end{minipage}
  \hfill
\end{figure}

\begin{figure}[p]
  \hfill
    \begin{minipage}[t]{.45\textwidth}
    \begin{center}
\includegraphics[width=0.9\textwidth]{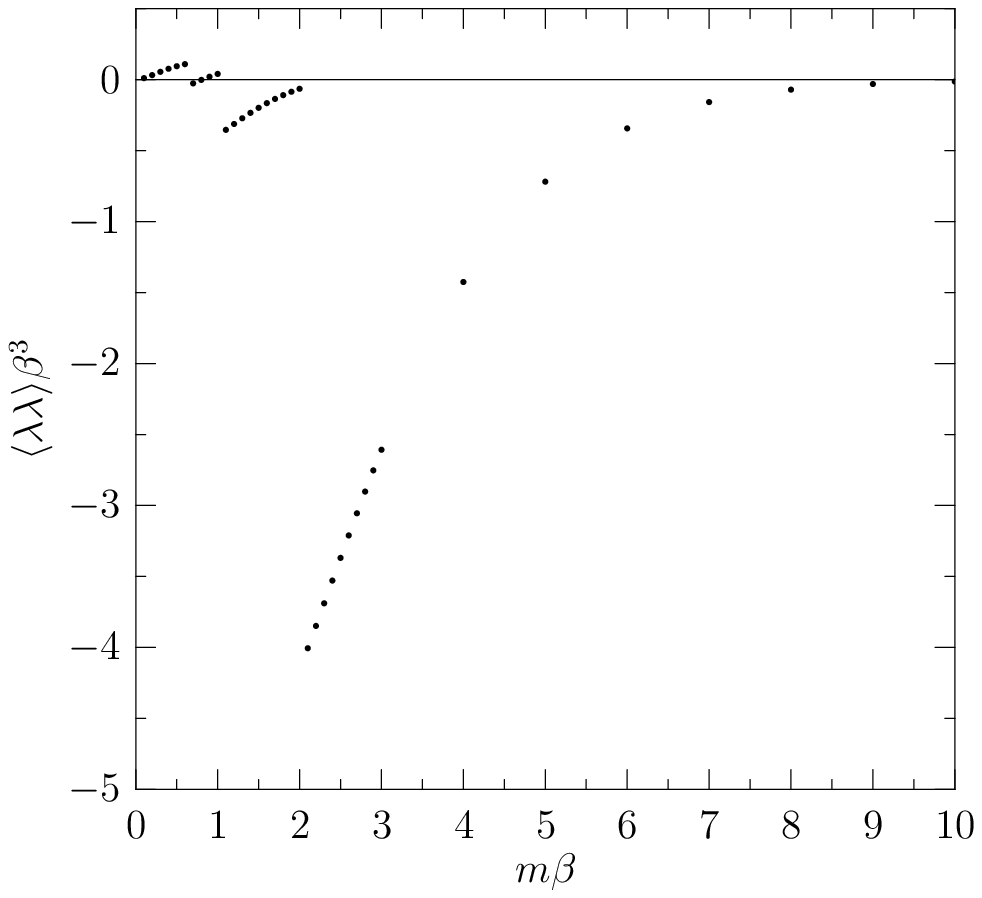}
       \caption{$\langle \lambda \lambda \rangle_{ADJ(+)}$ for $N = 6$, $N_f = 2$}
       \label{chi_adj_nc6_nf2}
    \end{center}
  \end{minipage}
  \hfill
  \begin{minipage}[t]{.45\textwidth}
    \begin{center}
\includegraphics[width=0.9\textwidth]{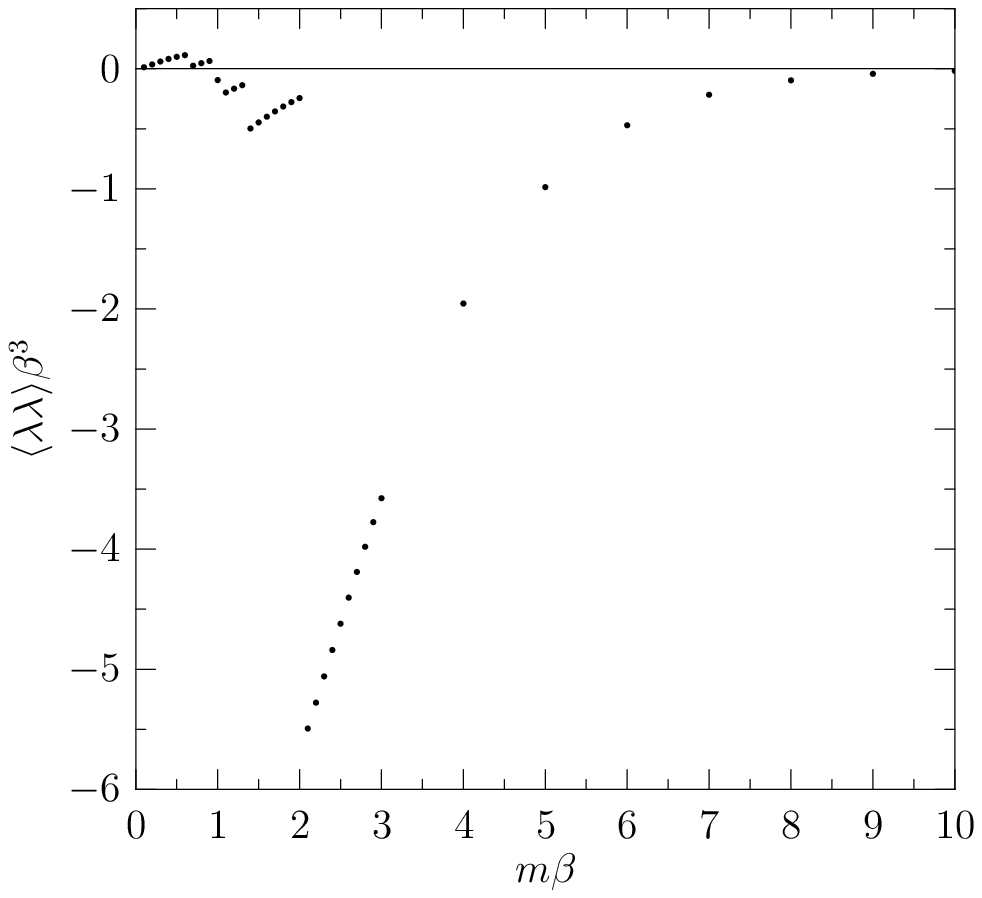}
\caption{$\langle \lambda \lambda \rangle_{ADJ(+)}$ for $N = 7$, $N_f = 2$}
\label{chi_adj_nc7_nf2}
    \end{center}
  \end{minipage}
  \hfill
\end{figure}

\begin{figure}[p]
  \hfill
    \begin{minipage}[t]{.45\textwidth}
    \begin{center}
\includegraphics[width=0.9\textwidth]{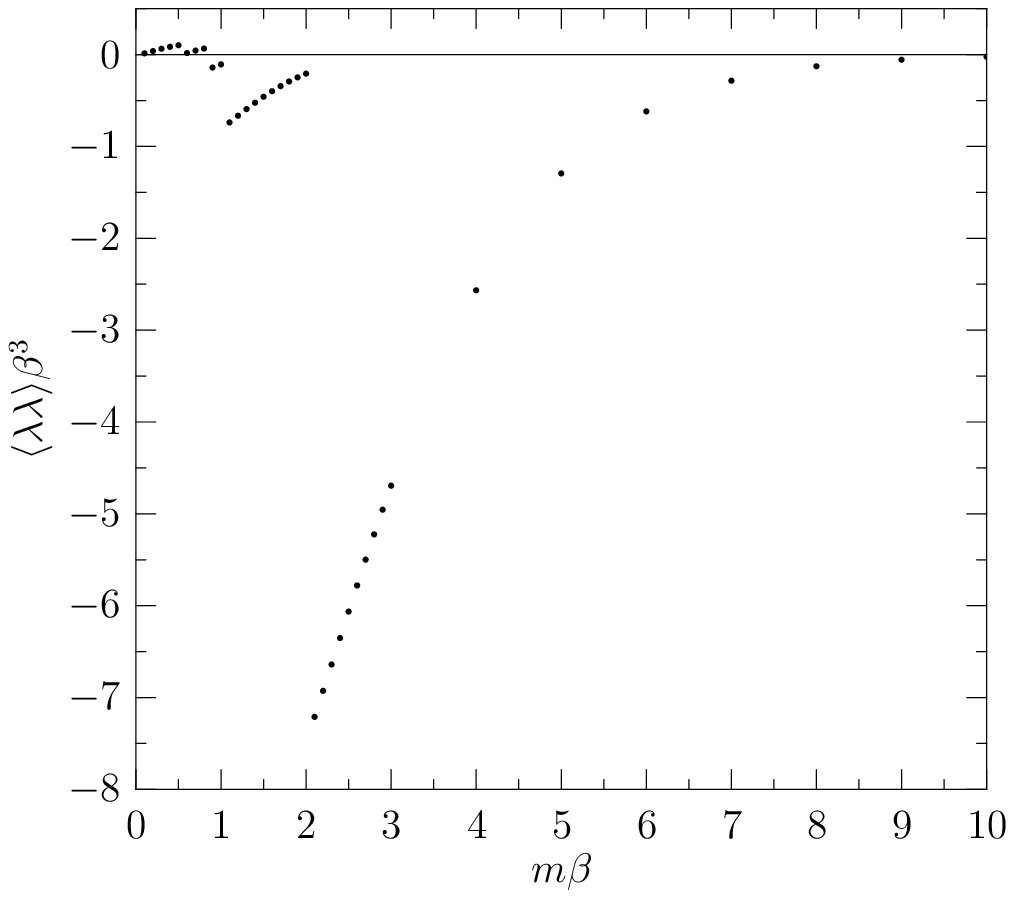}
       \caption{$\langle \lambda \lambda \rangle_{ADJ(+)}$ for $N = 8$, $N_f = 2$}
       \label{chi_adj_nc8_nf2}
    \end{center}
  \end{minipage}
  \hfill
  \begin{minipage}[t]{.45\textwidth}
    \begin{center}
\includegraphics[width=0.9\textwidth]{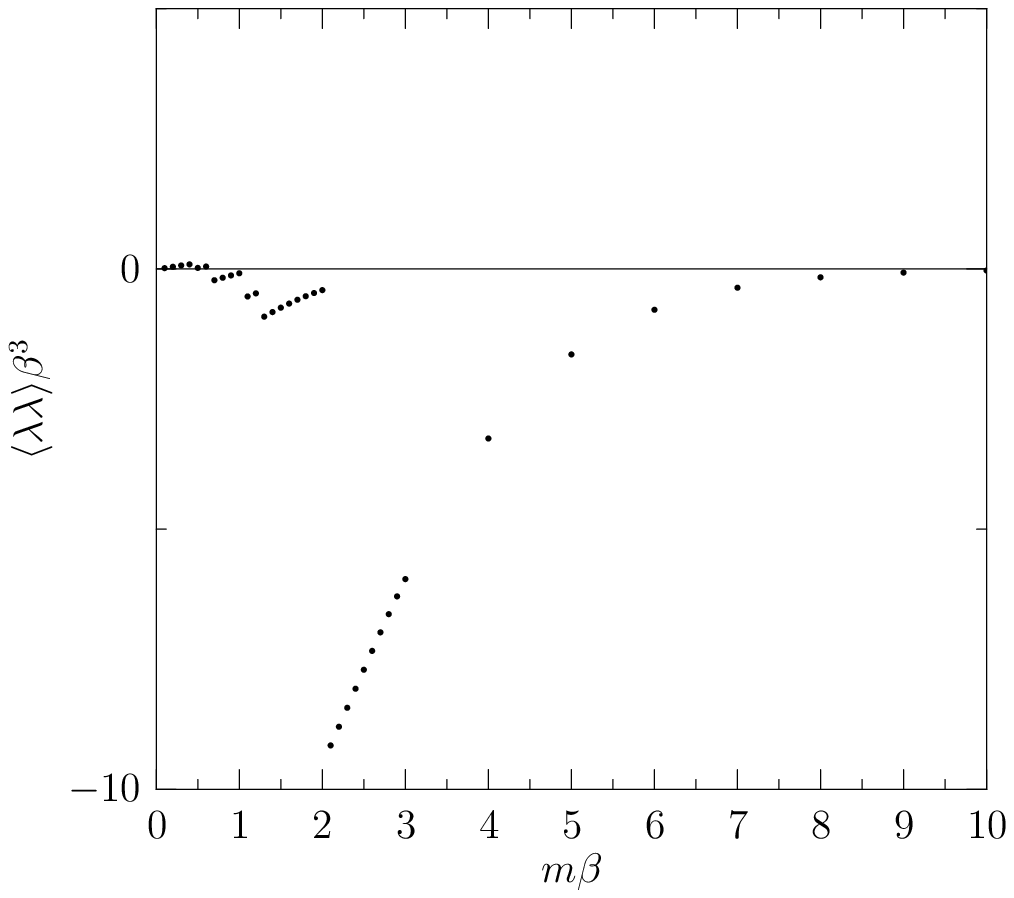}
\caption{$\langle \lambda \lambda \rangle_{ADJ(+)}$ for $N = 9$, $N_f = 2$}
\label{chi_adj_nc9_nf2}
    \end{center}
  \end{minipage}
  \hfill
\end{figure}

The chiral condensate in QCD(Adj) is given by

\begin{equation}
\langle \lambda \lambda \rangle (m) = - \lim_{V_4 \rightarrow \infty} \frac{1}{V_4 N_f} \frac{\partial}{\partial m} \ln Z (m) = \frac{1}{N_f} \frac{\partial}{\partial m} V_{eff} (P, m) .
\end{equation}

\noindent where we replaced ${\bar \psi} \psi$ in eq. (\ref{chicond}) with $\lambda \lambda$ since adjoint particles and antiparticles are indistinguishable. In order to ascertain the order of the transitions in our phase diagrams of $V_{Adj(+)}$ vs. $m \beta$ in Figures \ref{pt_adj_nc2_nf2} - \ref{pt_adj_nc9_nf2}, we calculated also the chiral condensates. These are shown for $N$ from $2$ through $9$ in Figures \ref{chi_adj_nc2_nf2} - \ref{chi_adj_nc9_nf2}. The most remarkable transition in all cases is that to the deconfined phase which exhibits the largest jump in $m \beta$. But, all phase transitions are marked by a noticeable jump in $\langle \lambda \lambda \rangle$ suggesting that all transitions are first order. There is also a smooth increase in $\langle \lambda \lambda \rangle$ at low $m \beta$ as observed in the case of fundamental fermions, and as well for symmetric and antisymmetric representation fermions.

\subsection{Orientifold and orbifold planar equivalence}

In 2003 - 2004, Armoni, Shifman, and Veneziano published a series of papers in which they proved non-perturbatively the equivalence of the bosonic sectors of QCD(Adj) with $N_f^M$ Majorana fermion flavours and QCD(AS/S) with $N_f^D = N_f^M$ Dirac fermion flavours, in the planar (or large $N$) limit \cite{Armoni:2003gp,Armoni:2004ub}. This is called orientifold planar equivalence. For $N = 3$, QCD(AS) is equivalent to QCD (taking $\mu \rightarrow - \mu$). Also, QCD(Adj) with one Majorana flavour of massless fermions is ${\cal N} = 1$ supersymmetric Yang-Mills (SYM) theory. This makes orientifold planar equivalence particularly useful since it allows one to use results in supersymmetry to make estimates in one-flavour QCD up to ${\cal O} (1 / N)$ corrections \cite{Armoni:2003fb}. For example, in \cite{Armoni:2003yv} the quark condensate was calculated in one-flavour QCD from the gluino condensate in SYM. This was subsequently verified in \cite{DeGrand:2006uy} using lattice simulations.


Perturbative comparison of the phase diagrams in QCD(Adj) and QCD(AS/S) was performed first in \cite{Sannino:2005sk}, then \cite{Unsal:2006pj}, then quite thoroughly in \cite{Unsal:2007fb}. In \cite{Unsal:2006pj} it is argued that orientifold planar equivalence is only valid when charge conjugation (${\cal C}$) symmetry is not broken. In the example provided, they showed that ${\cal C}$-symmetry is broken in $U(N)$ QCD(AS/S) with periodic boundary conditions for massless fermions on $S^1 \times {\field R}^3$, for small $S^1$. While it seems to be true that ${\cal C}$-symmetry is required in general for planar equivalence, their example is a special case in which orientifold planar equivalence actually still holds, despite ${\cal C}$-parity being broken. From the derivation of the fermion determinant in Appendix \ref{lnZ}, it is clear that changing fermion boundary conditions from periodic to antiperiodic results in shifting the gauge field by $i \pi / \beta$, i.e. $v \rightarrow v + \pi$. For the $U(N)$ theory (which corresponds to the $SU(N)$ theory for $N = 4, 8, 12, ...$) transforming from antiperiodic to periodic boundary conditions results in the the effective potential being minimized when the individual Polyakov loop angles are shifted each according to $v_i \rightarrow v_i + \pi / 2$. In terms of the partition function,

\begin{equation}
\begin{array}{lll}
QCD(AS/S,+):& \hspace{1cm} \ln Z_{AS/S,+} \sim g (v_i + v_j), &\hspace{1cm} v_i = \pi / 2 ,\\
QCD(AS/S,-):& \hspace{1cm} \ln Z_{AS/S,-} \sim g (v_i + v_j + \pi), &\hspace{1cm} v_i = 0 .
\end{array}
\end{equation}

\noindent Therefore, for the $U(N)$ QCD(AS/S) theories the partition function will always be the same under transformation of fermion boundary conditions between periodic and antiperiodic.

\begin{figure}[t]
  \hfill
  \begin{minipage}[t]{.45\textwidth}
    \begin{center}  
      \includegraphics[width=0.95\textwidth]{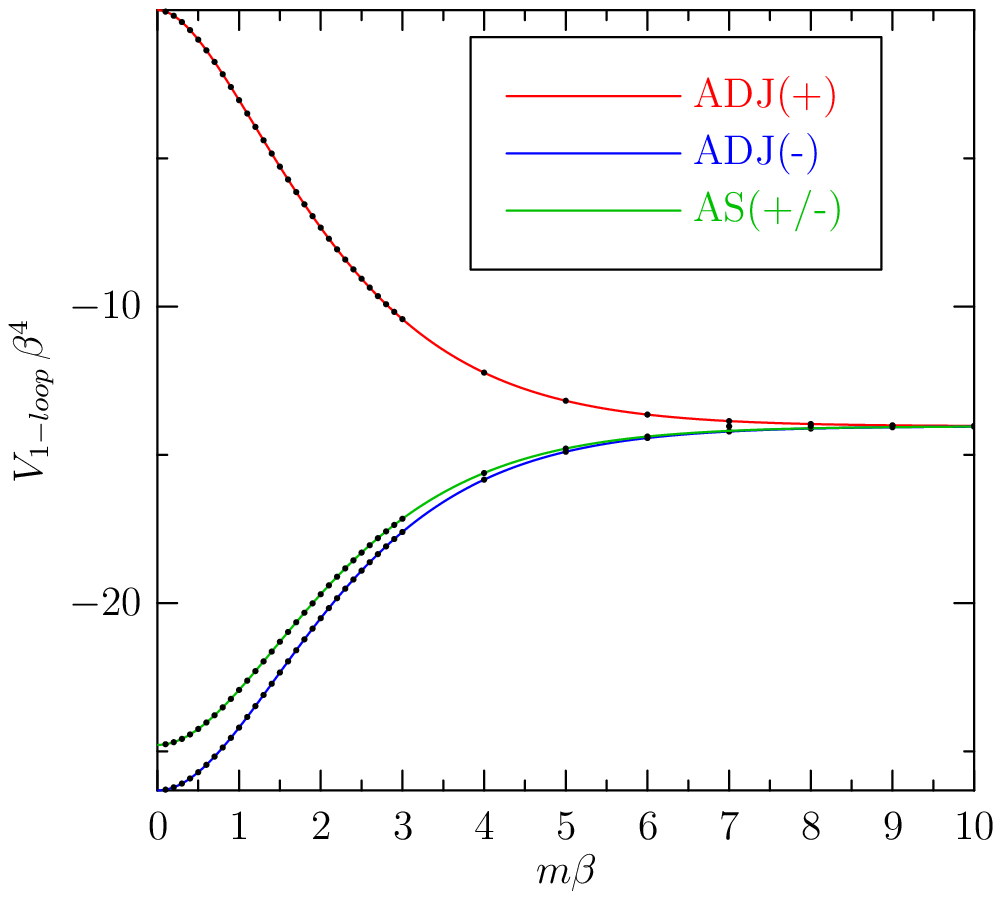}
    \end{center}
  \end{minipage}
  \hfill
  \begin{minipage}[t]{.45\textwidth}
    \begin{center}
\includegraphics[width=0.94\textwidth]{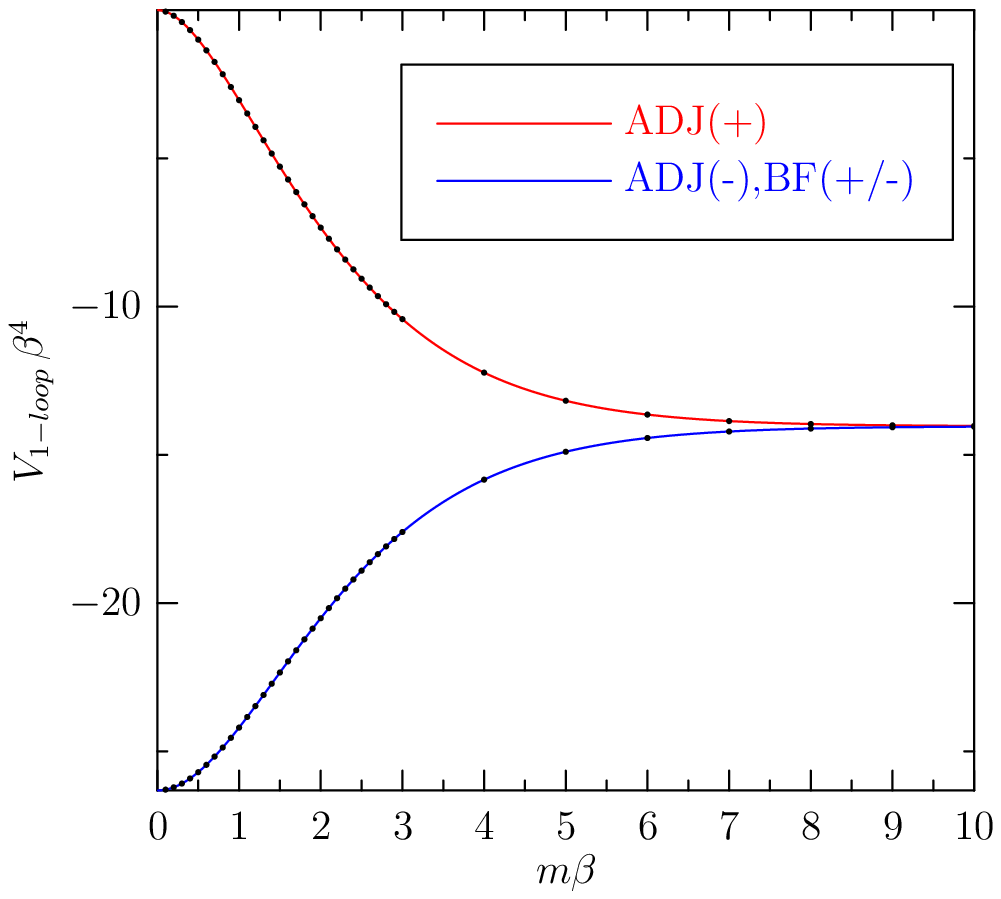}
    \end{center}
  \end{minipage}
  \hfill
  \caption{(Left) Orientifold planar equivalence: To check at one-loop that the equivalence is valid for QCD(Adj,-), QCD(AS,+), QCD(AS,-), we plot the effective potential in all three cases for the $U(N)$ theories at $N = 8$. For QCD(Adj) $N_f^M = 1$; for QCD(AS), $N_f^D = 1$. Comparison with QCD(Adj,+) indicates that this theory is not equivalent to any of the others in this limit. The effective potential for QCD(Adj,-) is minimized in the deconfined phase where the Polyakov loop angles are $v_i = 0 \,\, \forall i$. This is also true for QCD(AS,-). For QCD(AS,+) $V_{eff}$ is minimized in the ${\cal C}$-breaking phase where $v_i = \pi / 2 \,\, \forall i$. Regardless, the effective potentials of QCD(AS,+/-) correspond exactly. (Right) Orbifold equivalence: Analogously with the orientifold result we find at one-loop that orbifold equivalence is valid for QCD(Adj,-), QCD(BF,+), and QCD(BF,-), but not QCD(Adj,+). We plot the effective potentials $V_{Adj(\pm)}$ with $N_f^M = 1$, $\frac{1}{2} V_{BF(\pm)}$ with $N_f^D = 1$ for the $U(N)$ theories with $N = 8$. In QCD(BF,-) the effective potential is minimized in the deconfined phase with $v_{1,i} = v_{2,i} = 0 \,\, \forall i$. For QCD(BF,+) $V_{eff}$ is minimized in the $Z_2$ interchange broken phase where $v_{1,i} = \pi / 2$ and $v_{2,i} = - \pi / 2$, $\forall i$. However, the effective potentials $V_{BF,\pm}$ correspond exactly, and are identical to $2 V_{Adj,-}$.}
  \label{equivalence}
\end{figure}

In summary the equivalence holds in the perturbative limit when considering QCD(Adj,-), QCD(AS,+), and QCD(AS,-), but not QCD(Adj,+). This is shown explicitly in Figure \ref{equivalence} (Left) where the one-loop effective potentials of these theories in $U(N)$ are plotted together for $N = 8$. QCD(AS,+) and QCD(AS,-) correspond exactly. QCD(AS,+/-) and ACD(Adj,-) correspond up to ${\cal O} (1/N)$ corrections. These statements of course also hold for QCD(S) in place of QCD(AS). For the $SU(N)$ theories, the partition functions for QCD(AS/S) agree exactly with the $U(N)$ case when $N$ is a multiple of $4$, otherwise they agree up to ${\cal O} (1/N)$ corrections, as seen in eqs. (\ref{qcd_as_mult4}), (\ref{qcd_as_nodd}), (\ref{qcd_as_nminus2mult4}). In QCD(Adj), the $SU(N)$ theory has one less degree of freedom corresponding to an extra $-1$ in $\Tr_A P = \left| \Tr_F P \right|^2 - 1$ (see Appendix \ref{reps}), resulting in an extra constant in $g_{Adj,\pm}$ in the partition function [see eqs. (\ref{g_adj_pbc}) and (\ref{g_adj_abc})]. This constant is subleading in the large $N$ limit. Therefore, the above statements about orientifold planar equivalence hold in both the $U(N)$ and the $SU(N)$ theory, as expected.

Conclusions analogous to those for orientifold planar equivalence can also be shown to hold in a related orbifold planar equivalence as well. In \cite{Kovtun:2003hr} Kovtun, Unsal and Yaffe proved non-perturbatively the large-N orbifold equivalence of [$U(2 N)$] QCD(Adj) with $N_f^M$ Majorana fermion flavours and QCD(BF) [$U(N)_1 \times U(N)_2$ gauge theory with bifundamental representation fermions] with $N_f^D = N_f^M$ Dirac fermion flavours, in the large mass, strong coupling phase. This result was anticipated in \cite{Schmaltz:1998bg,Armoni:1999gc}. The technique of orbifold projection is quite useful and has been of recent interest as a means of testing volume independence in confining gauge theories via Eguchi-Kawai reduction \cite{Kovtun:2007py}.

In \cite{Tong:2002vp,Armoni:2005wta,Kovtun:2005kh} the breaking (or lack thereof) of $Z_2$ interchange symmetry of the two gauge fields is explored, the symmetry which is thought to be required for orbifold planar equivalence to hold. In \cite{Tong:2002vp} this $Z(2)$-symmetry was shown to be broken in QCD(BF,+) with a compact dimension. In \cite{Armoni:2005wta} it was argued that this symmetry is broken on $R^4$ when periodic boundary conditions are applied to bifundamental representation fermions, causing the equivalence to fail. In \cite{Kovtun:2005kh} it was argued that there is no evidence to support $Z(2)$-breaking on $R^4$.

As with the orientifold case, it is straightforward to show perturbatively that the effect of symmetry breaking resulting from the application of PBC on bifundamental fermions does not change the partition functions from that of the theory with antiperiodic boundary conditions on fermions \footnote{We would like to thank Mithat Unsal for pointing out this important calculation to us.}. The effect of $Z(2)$-symmetry breaking in QCD(BF,+) is cancelled out by the application of PBC on fermions so the equivalence does not fail. But, if the $Z(2)$ symmetry is broken by some means other than application of PBC on fermions, such that the partition function is different, then the equivalence would not hold.

QCD(BF) is a $U(N)_1 \times U(N)_2$ theory with gauge fields in the adjoint representation of each group, and bifundamental representation fermions which transform as the fundamental in the first group, and antifundamental in the second. The additional gauge field results in a doubling of the gauge boson term, which results from the action

\begin{equation}
S_{QCD(BF)} = \frac{1}{4 g^2} \int_0^{\beta} {\mathrm d} \tau \int {\mathrm d}^3 x \Tr_F \left( F_{1, \mu \nu} F_{1, \mu \nu} + F_{2, \mu \nu} F_{2, \mu \nu} \right) + {\bar \psi} \left( \Dsl_{BF} - \gamma_0 {\cal M} + M \right) \psi .
\end{equation}

\noindent $D^{\mu}_{BF} = \partial^{\mu} + A^{\mu}$ with $A^{\mu} = A_1^{\mu} + A_2^{\mu}$ where $A_1$ transforms as the fundamental and $A_2$ transforms as the antifundamental so the Polyakov loop is $P = e^{i v}$ with $v = v_1 + v_2$. The fermion contribution to the effective potential follows from eq. (\ref{fermion_det}) for the determinant with fermions in representation $R = BF$ so $\Tr_{BF} P = \Tr_F P_1 \Tr_F P_2^{\dagger}$ and $P_j = e^{i v_j} = {\text diag} \{ e^{i v_{j,1}}, ... , e^{i v_{j,N}} \}$, $j = 1, 2$. The effective potential (for $\mu = 0$) has the form

\begin{equation}
\begin{aligned}
V_{BF,\pm} = & - \frac{1}{\beta V_3} \ln \left[ \frac{{\rm det}^{N_f^D} \left[ - D_{BF,\pm}^2 (A) \right]}{{\rm det} \left[ - D_{A}^2 (A_1) \right] {\rm det} \left[ - D_{A}^2 (A_2) \right]} \right]\\
= & \, \frac{2 m^4 N_f}{3 \pi^2} \int_1^{\infty} {\mathrm d}t (t^2 - 1)^{3/2} {\rm Re} \left[ g_{BF,\pm} \left( \beta, m t, 0, v \right) \right]\\
& - \frac{2}{\pi^2 \beta^4} \sum_{n=1}^{\infty} \frac{1}{n^4} \left[ \Tr_A ( e^{i n v_1} ) + \Tr_A ( e^{i n v_2} ) \right] ,
\end{aligned}
\label{Vbf}
\end{equation}

\noindent where

\begin{equation}
g_{BF,+} = Tr_{BF} g(\beta, mt, 0, v) = \sum_{j, k} g (\beta, mt, 0, v_{1,j} - v_{2,k}) ,
\end{equation}

\begin{equation}
g_{BF,-} = Tr_{BF} g(\beta, mt, 0, v + \pi) = \sum_{j, k} g (\beta, mt, 0, v_{1,j} - v_{2,k} + \pi) .
\end{equation}

We minimize the effective potential in eq. (\ref{Vbf}) with respect to the $v_{1,j}$ and $v_{2,k}$ and obtain the expected results. With antiperiodic (-) boundary conditions on fermions the deconfined phase is always favoured:

\begin{equation}
v_{1,j} = v_{2,j} = 0 , \hspace{1cm} \forall \,\, j = 1, ... , N.
\end{equation}

\noindent When periodic (+) boundary conditions are applied to bifundamental fermions

\begin{equation}
v_{1,j} = \frac{\pi}{2} , \hspace{1cm} v_{2,j} = - \frac{\pi}{2} , \hspace{1cm} \forall \,\, j = 1, ... , N ,
\end{equation}

\noindent or any rotation, of all the eigenvalues, such that $v_{1,j} - v_{2,k}$ remains an odd multiple of $\pi$, in agreement with \cite{Tong:2002vp,Unsal:2007fb}. Here the $Z_2$ exchange symmetry between the gauge field eigenvalue angles $v_{1,j}$ and $v_{2,k}$ is clearly broken. However, the effective potentials are the same regardless of the fermion boundary conditions. Again we look at the form of the partition functions:

\begin{equation}
\begin{array}{lll}
QCD(BF,+):& \hspace{1cm} \ln Z_{BF,+} \sim g (v_{1,j} - v_{2,k}), &\hspace{1cm} v_{1,j} = \pi / 2 , \,\, v_{2,k} = - \pi / 2\\
QCD(BF,-):& \hspace{1cm} \ln Z_{BF,-} \sim g (v_{1,j} - v_{2,k} + \pi), &\hspace{1cm} v_{1,j} = v_{2,k} = 0 .
\end{array}
\end{equation}

\noindent This is shown explicitly for $N = 8$ in Figure \ref{equivalence} (Right) where comparison is made with QCD(Adj,+/-). The conclusion in terms of the effective potentials is that $\frac{1}{2} V_{BF,\pm} = V_{Adj,-}$ [where we considered $U(N)$ QCD(Adj) for the purpose of comparing with the orientifold equivalence. For $U(2 N)$ QCD(Adj) $V_{Adj,-} = 2 V_{BF,\pm}$] but there is no equivalence with $V_{Adj,+}$ (at least not at one loop). The factor of $1 / 2$ results because of the additional gauge boson term in QCD(BF) and the fact that the number of {\it Dirac} flavours multiplies the fermion term, where the same number of Majorana flavours (one Dirac flavour is two Majorana) multiplies the fermion term in QCD(Adj). Of course these conclusions also hold for the $SU(N)$ theories in the large $N$ limit.

\section{Conclusions}

We can summarize the results and draw some conclusions concerning one-loop calculations of the phase diagrams of QCD(R) for fermions in the fundamental, antisymmetric, symmetric, and adjoint representations. We have minimized the one loop effective potential with respect to the Polyakov loop eigenvalues, for a range of $m \beta$, on the topology $S^1 \times {\field R}^3$, for small $S^1$. For ABC on fermions the deconfined phase is favoured for all $m \beta$ in $SU(N)$ gauge theories with all the above fermion representations. With PBC on fundamental fermions a phase in which the Polyakov loop eigenvalue angles are all close to $\pi$ is preferred. For $N$ odd this results in ${\cal C}$-symmetry breaking since the eigenvalue angles only approach $\pi$, including an extra ${\cal O} (1/N)$ contribution. For PBC on antisymmetric/symmetric representation fermions a ${\cal C}$-breaking phase is always favoured where the Polyakov loop angles prefer to be $\pm \pi / 2$ [including additional ${\cal O} (1/N)$ contributions for $N$ not a multiple of $4$]. For PBC on adjoint representation fermions and $N_f = 1$ Majorana fermion flavour the deconfined phase is favoured for all $m \beta > 0$. However, for $N_f \ge 2$ the phase structure is quite varied and depends on $N$.

Concerning orientifold planar equivalence, if ${\cal C}$-symmetry is broken in QCD(AS/S) by applying periodic boundary conditions to fermions, then the equivalence still holds because the effect of transforming boundary conditions and of ${\cal C}$-parity breaking cancel each other out, causing the partition function to be the same [for the $U(N)$ theory, and with ${\cal O} (1/N)$ corrections for the $SU(N)$ theory] as when antiperiodic fermion boundary conditions are applied. However, if ${\cal C}$-symmetry is broken in some other way, such that the effective potential is not minimized to the same value as for QCD(AS/S,-) without ${\cal C}$-parity breaking, then the equivalence fails.

Considering orbifold equivalence, the breaking of $Z_2$ exchange symmetry when periodic boundary conditions are applied to bifundamental representation fermions results in the same partition function as when antiperiodic boundary conditions are applied. Therefore the equivalence still holds because the effect of the changing fermion boundary conditions from antiperiodic to periodic adds a factor of $\pi$ to the gauge fields that cancels out the $Z_2$ exchange symmetry breaking.

From our results, and from the derivation of the partition function in Appendix \ref{lnZ}, several well-established group theory equivalences are clear. For example, QCD(S) is the same as QCD(Adj) in $SU(2)$ because the symmetric representation is the same as the adjoint (see Appendix \ref{reps}). QCD(AS) is the same as QCD(F) in $SU(3)$ if $\mu \rightarrow - \mu$ because the antisymmetric and the antifundamental representations correspond.

Considering the partition function we also have, in $U(N)$ [or $SU(N)$ with N = 4, 8, 12, ...], $Z_{AS,-}$ the same as $Z_{AS,+}$, even though the Polyakov loop eigenvalues differ. Of course, this is also true for $Z_{S,-}$ and $Z_{S,+}$. The same is true for $Z_{BF,-}$ and $Z_{BF,+}$ which are identical considering the $U(N)$ theories [and agree to ${\cal O} (1/N)$ corrections considering the $SU(N)$ theories] and correspond to $Z_{AS/S,\pm}$ in the large-$N$ limit. Whether these equivalences hold non-perturbatively is not clear.

The results in this paper were calculated the case of $\mu = 0$, however it is straightforward to include a finite chemical potential using the formulas in Appendix \ref{lnZ}. The only restriction is that the integral representation is only valid for $\mu < m$, but the series representation often converges rather quickly.

\section{Acknowledgments}

We would like to thank Adi Armoni, Agostino Patella, Ben Svetitsky, Mithat Unsal, and Lawrence Yaffe for useful discussions. We would also like to thank the Institute for Nuclear Theory for their hospitality during the "New frontiers in large N gauge theories" conference where this research was discussed and progress was made towards its completion.

\clearpage
\appendix
\begin{section}{\label{lnZ}Derivation of $\ln Z_{QCD(R)}$}

For $N_f$ quark flavours in representation $R$, with masses $m_f$, at chemical potential $\mu_f$, and at inverse temperature $\beta = 1/T$, the QCD(R) partition function is given by the (Euclidean space) path integral

\begin{equation}
Z_{QCD(R)} \left( \beta, \mu \right) = \int {\mathrm {\cal D}}A \, e^{-S_{YM} (A)} \int {\mathrm {\cal D}} {\bar \psi} {\mathrm {\cal D}} \psi e^{-\int_0^{\beta} {\mathrm d} \tau \int {\mathrm d}^3 {\bf x} \, {\bar \psi} \left( \Dsl_R - \gamma_0 {\cal M} + M \right) \psi} ,
\label{Zqcd}
\end{equation}

\noindent where the gauge field $A_{\mu} (x)= T_R^a A_{\mu}^a (x)$, $\mu = 0, 1, 2, 3$, the $a$ indices are $a = 0, ..., N^2 - 1$, and the $T_R^a$ are the generators of $SU(N)$ in the representation $R$. $\psi$ is a vector of fermion fields containing $N_f$ anti-commuting 4-spinors $\psi_f$ in representation R. ${\bar \psi}$ contains the corresponding antifermion fields. $M$ is the fermion mass matrix where $(M)_{f f'} = m_f \delta_{f f'}$ and ${\cal M}$ is the fermion chemical potential matrix where $({\cal M})_{f f'} = \mu_f \delta_{f f'}$. $\Dsl_R = \gamma_{\mu} D_{\mu}$ where $D_{\mu}$ is the covariant derivative

\begin{equation}
D_{\mu} = \partial_{\mu} + A_{\mu} .
\label{cov_deriv}
\end{equation}

\noindent The $A_{\mu}$ are $D(R) \times D(R)$ matrices in the colour space, taken in the representation R of the fermion fields, and $\gamma_{\mu}$ are the Euclidean space Hermitian gamma matrices satisfying $\{ \gamma_{\mu}, \gamma_{\nu} \} = 2 \delta_{\mu \nu}$. We work in Euclidean space throughout this paper with the metric $g_{\mu \nu} = \delta_{\mu \nu}$. Note that in eq. (\ref{cov_deriv}) we rescaled $A$ to absorb the coupling $g$ according to $g A_{\mu} \rightarrow A_{\mu}$.

$S_{YM} (A)$ is the pure Yang-Mills theory action which contains only the boson contribution. It is given by

\begin{equation}
S_{YM} = \frac{1}{4 g^2} \int_0^{\beta} {\mathrm d} \tau \int {\mathrm d}^3 x \, \Tr_F \left( F_{\mu \nu} F_{\mu \nu} \right) ,
\end{equation}

\noindent where $F_{\mu \nu}$ is the field strength defined by

\begin{equation}
F_{\mu \nu} \equiv \left[ D_{\mu}, D_{\nu} \right] = \partial_{\mu} A_{\nu} - \partial_{\nu} A_{\mu} + \left[ A_{\mu}, A_{\nu} \right] .
\end{equation}

Because the functional integrals in eq. (\ref{Zqcd}) have an exponential that is quadratic in the fermion fields $\psi_f$ they are of the gaussian form and exactly solvable. The result is

\begin{equation}
Z_{QCD(R)} \left( \beta, \mu \right) = \int {\mathrm {\cal D}}A \, \det \left( \Dsl_R - \gamma_0 {\cal M} + M \right) e^{-S_{YM} (A)} .
\label{det_Zqcd}
\end{equation}

We need the eigenvalues of $\Dsl_R$ to evaluate the determinant for fermion fields $\psi$ in representation $R$. For free fields this is more easily done in momentum space using

\begin{equation}
\ln Z = \beta V_3 \int \frac{{\mathrm d}^4 p}{\left( 2 \pi \right)^4} \ln Z (p) ,
\end{equation}

\noindent where now $\Dsl_R$ in eq. (\ref{det_Zqcd}) is evaluated by having it act on the fermion fields in momentum space. These have the form

\begin{equation}
\psi \left( \tau, {\bf x} \right) = \frac{1}{\beta} \sum_{n \in {\field Z}} \int \frac{{\mathrm d}^3 p}{(2 \pi)^3} e^{i \left( {\bf p} \cdot {\bf x} + \omega_n^{\pm} \tau \right)} {\tilde \psi} \left( \omega_n^{\pm}, {\bf p} \right) ,
\label{ft_psi}
\end{equation}

\noindent where $V_3$ is the 3-volume, and $\omega_n^{\pm}$ are the Matsubara frequencies for periodic (+) or antiperiodic (-) boundary conditions applied to fermions,

\begin{equation}
\begin{aligned}
\omega_n^{+} &= 2 n \pi / \beta ,\\
\omega_n^{-} &= (2 n + 1) \pi / \beta .
\end{aligned}
\end{equation}

The compactification of $S^1$ also leads to periodicity in the gauge field $A$. In this case the boundary conditions are periodic (+) and we expand $A_{\mu}$ in a Fourier series according to

\begin{equation}
A_{\mu} \left( \tau, {\bf x} \right) = \frac{1}{\beta} \sum_{n \in {\field Z}} e^{i \omega_n^{+} \tau} A_{\mu} \left( \omega_n^{+}, {\bf x} \right) .
\end{equation}

Using eq. (\ref{cov_deriv}) for $D_{\mu}$ and eq. (\ref{ft_psi}) for the eigenvectors $\psi$ we can compute the eigenvalues $i \lambda$ of $\Dsl_R$ in frequency-momentum space. The result is

\begin{equation}
i \lambda = i \gamma_0 \omega_n^{\pm} + i \gamma \cdot {\bf p} + \gamma_{\mu} A_{\mu}^a T^a ,
\end{equation}

\noindent where $i \lambda$ is a constant times the identity matrix in colour ($k = 1, ..., N$), flavour ($f = 1, ..., N_f$), and spinor space. $A_{\mu}$ are elements of the Lie algebra of $SU(N)$. Here we have chosen $A_{\mu}$ to be $N \times N$ and transform as the representation $R$. Then the $T^a$ are the generators of $SU(N)$ in the fundamental representation.

$\Dsl_R$ is anti-Hermitian ($\Dsl_R^{\dagger} = -\Dsl_R$) since the gamma matrices are Hermitian, $\omega_n$ and ${\bf p}$ are real, and the $A_{\mu}$ are anti-Hermitian. Therefore, the eigenvalues $i \lambda$ of $\Dsl_R$ are pure imaginary and $\lambda \in \Re$. Also, since the Dirac operator satisfies $\{ \gamma_5, \Dsl_R \} = 0$, and since $\{ \gamma_0, \gamma_5 \} = 0$, the eigenvalues of $\Dsl_R - \gamma_0 \mu_f$ show up in pairs in frequency-momentum space. The eigenvalue equations are

\begin{equation}
\begin{aligned}
\left( \Dsl_R - \gamma_0 {\cal M} \right) \psi &= \left( i \lambda - \gamma_0 {\cal M} \right) \psi ,\\
\left( \Dsl_R - \gamma_0 {\cal M} \right) \left( \gamma_5 \psi \right) &= - \left( i \lambda - \gamma_0 {\cal M} \right) \left( \gamma_5 \psi \right) ,
\end{aligned}
\end{equation}

\noindent which agrees with \cite{Verbaarschot:2000dy, Dalmazi:2002uh} for the fundamental representation with ${\cal M} = 0$.

The fermion determinant in eq. (\ref{det_Zqcd}) is over the spinor indices, colour indices, flavour indices, as well as over frequency-momentum space. We first take the determinant over the colour indices ($k = 1, ..., N$) using the identity, $\ln \det A = \Tr \ln A$, for an aribtrary $N \times N$ matrix $A$. Then we take the determinant over the spinor indices ($4$ degrees of freedom for Dirac spinors and $2$ for Majorana spinors), then frequency-momentum space indices, then finally over flavour indices. This proceeds as

\begin{equation}
\begin{aligned}
&\ln \det \left( \Dsl_R - \gamma_0 {\cal M} + M \right)\\
&= \Tr_R \ln \det \left( \Dsl_{R} - \gamma_0 {\cal M} + M \right)\\
&= 2 \, \Tr_R \sum_n \sum_{{\bf p}} \ln \det \left[ \left( i \lambda - \gamma_0 {\cal M} + M \right) \left( -i \lambda + \gamma_0 {\cal M} + M \right) \right]\\
&= 2 \, \Tr_R \sum_n \sum_{{\bf p}} \ln \det \left[ - \left( i \lambda - \gamma_0 {\cal M} \right)^2 + M^2 \right]\\
&= 2 \, \Tr_R \sum_n \sum_{{\bf p}} \ln \det \left[ - \left( i \omega_n^{\pm} + A_{0} - {\cal M} \right)^2 - \left( i {\bf p} + {\bf A} \right)^2 + M^2 \right]\\
&= 2 \beta V_3 \sum_{f=1}^{N_f} \, \Tr_R \int \frac{{\mathrm d}^4 p}{(2 \pi)^4} \ln \left[ \left( \omega_n^{\pm} - i A_{0} + i \mu_f \right)^2 + \left( {\bf p} - i {\bf A} \right)^2 + m_f^2 \right] .
\end{aligned}
\end{equation}

\noindent To get from line 4 to line 5 we used, in $\lambda$, the Euclidean gamma matrices $\gamma_{\mu}$ in the chiral representation where $\gamma_5 \equiv \gamma_1 \gamma_2 \gamma_3 \gamma_0 = \text{diag} \{ 1, 1, -1, -1 \}$. To see clearly the effect of antiparticles we can rearrange the result as follows

\begin{equation}
\begin{aligned}
&2 \sum_{n \in {\field Z}} \ln \left[ \left( \omega_n^{\pm} - i A_{0} + i \mu_f \right)^2 + \left( {\bf p} - i {\bf A} \right)^2 + m_f^2 \right]\\
&= \sum_{n \in {\field Z}} \left( \ln \left[ (\omega_n^{\pm})^{2} + \left( \omega_f - \left( \mu_f - A_0 \right) \right)^2 \right] + \ln \left[ (\omega_n^{\pm})^{2} + \left( \omega_f + \left( \mu_f - A_0 \right) \right)^2 \right] \right) ,
\end{aligned}
\label{antipart}
\end{equation}

\noindent where $\omega_f \equiv \sqrt{ \left( {\bf p} - i {\bf A} \right)^2 + m_f^2}$. The second term in (\ref{antipart}) is the antiparticle contribution which has $A_0 \rightarrow A_0^{\dagger}$ and $\mu_f \rightarrow - \mu_f$ with respect to the particle contribution as expected. To simplify later calculations we now take all the fermion masses and chemical potentials to be degenerate $m_f = m$ ($\omega_f = \omega$), $\mu_f = \mu, \forall f$.

The final result for the fermion contribution is then

\begin{equation}
\begin{aligned}
&\ln \det \left( \Dsl_R - \gamma_0 {\cal M} + M \right)\\
&= V_3 N_f \, \Tr_R \sum_{n \in {\field Z}} \int \frac{{\mathrm d}^3 p}{(2 \pi)^3} \left[ \ln \left( (\omega_n^{\pm})^{2} + \left( \omega - u \right)^2 \right) + \ln \left( (\omega_n^{\pm})^{2} + \left( \omega + u \right)^2 \right) \right] ,
\end{aligned}
\end{equation}

\noindent where $u = \mu - A_0$. Defining the zero-temperature determinant to 1 and evaluating as in \cite{Kapusta:2006pm} we get

\begin{equation}
\begin{aligned}
&\ln \det \left( \Dsl_R - \gamma_0 {\cal M} + M \right)\\
&= 2 V_3 N_f \, \Tr_R \int \frac{{\mathrm d}^3 p}{(2 \pi)^3} \left[ \ln \left( 1 \mp e^{- \beta \left( \omega - u \right)} \right) + \ln \left( 1 \mp e^{- \beta \left( \omega + u \right)} \right) \right] .
\end{aligned}
\end{equation}

\noindent where the top sign corresponds to periodic boundary conditions [PBC (+)] on fermions and the bottom sign is for the usual case of antiperiodic boundary conditions [ABC (-)] on fermions. Spherical symmetry in $p$ suggests evaluating the integral by converting to hyper-spherical coordinates using 

\begin{equation}
\Omega_d = \frac{2 \pi^{d/2}}{\Gamma(d/2)}
\end{equation}

\noindent and

\begin{equation}
\Gamma \left( n + \frac{1}{2} \right) = \frac{\left( 2 n \right) !}{n! 2^{2 n}} \sqrt{\pi} .
\end{equation}

\noindent Then

\begin{equation}
\begin{aligned}
&\ln \det \left( \Dsl_R - \gamma_0 {\cal M} + M \right)\\
& = 2 V_3 N_f \frac{\Omega_d}{(2 \pi)^d} \, \Tr_R \int_0^{\infty} {\mathrm d}p \, p^{d-1} \left[ \ln \left( 1 \mp e^{- \beta \left( \omega - u \right)} \right) + \ln \left( 1 \mp e^{- \beta \left( \omega + u \right)} \right) \right] .
\end{aligned}
\end{equation}

The Polyakov loop is defined as the path-ordered exponential of the temporal component of the gauge field,

\begin{equation}
P({\vec x}) = {\cal P} e^{\int_0^{\beta} {\mathrm d}t A_0 (x)} .
\end{equation}

\noindent For a constant background field defined by $A_0 \equiv i v / \beta$ field the Polyakov loop is

\begin{equation}
P = e^{\beta A_0} = e^{i v} ,
\end{equation}

\noindent where we have chosen a gauge in which $A_0$ is diagonal and $v$ is real, diagonal and traceless with elements $(v)_{i j} = v_i \delta_{i j}$. Then

\begin{equation}
P = {\rm diag} \{ e^{i v_1}, ... , e^{i v_N} \} .
\end{equation}

We make a further simplification by taking $A_i = 0$ for $i = 1, 2, 3$. Then $\omega = \sqrt{p^2 + m^2}$ and

\begin{equation}
\begin{aligned}
&\ln \det \left( \Dsl_R - \gamma_0 {\cal M} + M \right)\\
& = 2 V_d N_f \frac{\Omega_d}{(2 \pi)^d} \, \Tr_R \int_0^{\infty} {\mathrm d}p \, p^{d-1} \left[ \ln \left( 1 \mp e^{- \beta \left( \omega - \mu \right)} P^{\dagger} \right) + \ln \left( 1 \mp e^{- \beta \left( \omega + \mu \right)} P \right) \right]\\
& = - \frac{V_d N_f m^{(d+1)/2}}{2^{(d-3)/2} \pi^{(d+1)/2} \beta^{(d-1)/2}}\, \Tr_R \sum_{n=1}^{\infty} \frac{\left( \pm 1 \right)^n}{n^{(d+1)/2}} \left( e^{n \beta \mu} P^{\dagger n} + e^{- n \beta \mu} P^{n} \right) K_{(d+1)/2} (n \beta m) .
\end{aligned}
\end{equation}

We are interested in the physical case of $d=3$. Then using an integral representation for modified Bessel functions of the second kind:

\begin{equation}
K_{\nu} (z) = \frac{\sqrt{\pi} z^{\nu}}{2^{\nu} \Gamma (\nu + 1/2)} \int_1^{\infty} {\mathrm d}t e^{-z t} (t^2 - 1)^{\nu - 1/2} ,
\end{equation}

\noindent valid for $\nu > -1/2$ and ${\mathrm Re} (z) > 0$, the log of the fermion determinant becomes

\begin{equation}
\begin{aligned}
&\ln \det \left( \Dsl_R - \gamma_0 {\cal M} + M \right)\\
& = - \frac{V_3 N_f m^{2}}{\pi^{2} \beta}\, \sum_{n=1}^{\infty} \frac{(\pm 1)^n}{n^{2}} \left[ e^{n \beta \mu} \, \Tr_R ( P^{\dagger n} ) + e^{- n \beta \mu} \, \Tr_R (P^{n}) \right] K_{2} (n \beta m)\\
& = - \frac{m^4 N_f \beta V_3}{3 \pi^2} \int_1^{\infty} {\mathrm d}t (t^2 - 1)^{3/2} \left[ g_{R,\pm} \left( \beta, m t, \mu, v \right) + g_{R,\pm}^{\dagger} \left( \beta, m t, - \mu, v \right) \right] ,
\end{aligned}
\label{fermion_det}
\end{equation}

\noindent where $g_{R,\pm}$ depends on the group representation of the fermions and the boundary conditions. We define the matrix

\begin{equation}
\begin{aligned}
g( \beta, m t, \mu, v ) &\equiv \sum_{n=1}^{\infty} e^{- i n v - n \beta (m t - \mu)}\\
&= \frac{e^{- i v} - e^{-\beta (m t - \mu)}}{e^{\beta (m t - \mu)} - 2 \cos v + e^{- \beta (m t - \mu)}} ,
\end{aligned}
\end{equation}

\noindent where the first line converges to the second line for $\mu < m$. We study fermions in the fundamental (F), adjoint (Adj), symmetric (S), and antisymmetric (AS) representations. For periodic boundary conditions (+) $g_{R,+}$ is defined according to $g_{R,+} \equiv \Tr_R g ( \beta, m t, \mu, v )$. Then

\begin{equation}
g_{F,+} \equiv \Tr_F \, g ( \beta, m t, \mu, v ) = \sum_{i=1}^{N} g ( \beta, m t, \mu, v_i ) = \sum_{i=1}^{N} \frac{e^{- i v_i} - e^{-\beta (m t - \mu)}}{e^{\beta (m t - \mu)} - 2 \cos v_i + e^{- \beta (m t - \mu)}} ,
\label{gpbcf}
\end{equation}

\begin{equation}
\begin{aligned}
g_{Adj,+} &\equiv \Tr_{A} \, g ( \beta, m t, \mu, v ) = \sum_{i, j=1}^{N} g ( \beta, m t, \mu, v_i - v_j ) - g ( \beta, m t, \mu, 0 )\\
&= \sum_{i, j=1}^{N} \frac{e^{- i \left( v_i - v_j \right)} - e^{-\beta (m t - \mu)}}{e^{\beta (m t - \mu)} - 2 \cos \left( v_i - v_j \right) + e^{- \beta (m t - \mu)}} - \frac{1 - e^{-\beta (m t - \mu)}}{e^{\beta (m t - \mu)} - 2 + e^{- \beta (m t - \mu)}} ,
\end{aligned}
\label{g_adj_pbc}
\end{equation}

\begin{equation}
\begin{aligned}
g_{AS,+} &\equiv \Tr_{AS} \, g ( \beta, m t, \mu, v ) = \sum_{i<j=1}^{N} g ( \beta, m t, \mu, v_i + v_j)\\
&= \sum_{i<j=1}^{N} \frac{e^{- i ( v_i + v_j)} - e^{-\beta (m t - \mu)}}{e^{\beta (m t - \mu)} - 2 \cos ( v_i + v_j ) + e^{- \beta (m t - \mu)}} ,
\end{aligned}
\label{gpbcas}
\end{equation}

\begin{equation}
\begin{aligned}
g_{S,+} &\equiv \Tr_{S} \, g ( \beta, m t, \mu, v ) = \sum_{i \le j=1}^{N} g ( \beta, m t, \mu, v_i + v_j)\\
&= \sum_{i \le j=1}^{N} \frac{e^{- i ( v_i + v_j)} - e^{-\beta (m t - \mu)}}{e^{\beta (m t - \mu)} - 2 \cos ( v_i + v_j ) + e^{- \beta (m t - \mu)}} ,
\end{aligned}
\label{gpbcs}
\end{equation}

\noindent where the form of the higher dimensional representations can be derived in terms of the fundamental using the Frobenius formula in combination with tensor product formulae from Young tableaux as detailed in Appendix \ref{reps}. To get the results for antiperiodic boundary conditions applied to fermions we just take $v \rightarrow v + \pi$, so $g_{R,-} \equiv \Tr_R g ( \beta, m t, \mu, v + \pi )$. Then

\begin{equation}
g_{F,-} \equiv \sum_{i=1}^{N} g ( \beta, m t, \mu, v_i + \pi ) ,
\end{equation}

\begin{equation}
g_{Adj,-} \equiv \sum_{i, j=1}^{N} g ( \beta, m t, \mu, v_i - v_j + \pi ) - g ( \beta, m t, \mu, \pi ) ,
\label{g_adj_abc}
\end{equation}

\begin{equation}
g_{AS,-} \equiv \sum_{i<j=1}^{N} g ( \beta, m t, \mu, v_i + v_j + \pi ) ,
\end{equation}

\begin{equation}
g_{S,-} \equiv \sum_{i \le j=1}^{N} g ( \beta, m t, \mu, v_i + v_j + \pi ) .
\end{equation}

To obtain the phase diagram we need to minimize the free energy which is given by the effective potential:

\begin{equation}
\begin{aligned}
f = V_{eff} (m, \mu) &\equiv - \frac{1}{\beta} \frac{\partial \ln Z}{\partial V_3}\\
&= - \frac{1}{\beta V_3} \ln Z(m, \mu)\\
&= - \frac{1}{\beta V_3} \left[ \ln Z_B (m, \mu) + \ln Z_F (m, \mu) \right] ,
\end{aligned}
\end{equation}

\noindent where the fermions have been integrated out and their contribution to the effective potential is:

\begin{equation}
\begin{aligned}
V_{eff}^F (m, \mu) &= - \frac{1}{\beta V_3} \ln Z_F (m, \mu)\\
&= \frac{m^4 N_f}{3 \pi^2} \int_1^{\infty} {\mathrm d}t (t^2 - 1)^{3/2} \left[ g_{R,\pm} \left( \beta, m t, \mu, v \right) + g_{R,\pm}^{\dagger} \left( \beta, m t, - \mu, v \right) \right] .
\end{aligned}
\end{equation}

The one-loop contribution is calculated by introducing fluctuations ${\bar A}_{\mu}$ around a slowly varying background field $a_{\mu}$ such that $A_{\mu} = a_{\mu} + g {\bar A}_{\mu}$. The free energy has the form

\begin{equation}
f \sim T^4 \left[ c_0 + c_2 g^2 + c_3 g^3 + ... \right] ,
\end{equation}

\noindent where the one-loop result corresponds to the $c_0$ term. The contributing Feynman diagrams at one loop are shown in Figure \ref{oneloops}.

\begin{figure}[h]
\begin{center}
\includegraphics[width=10cm]{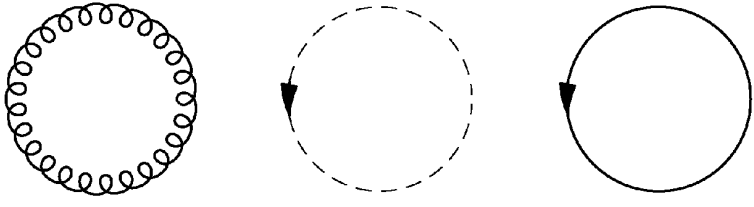}
\end{center}
\caption{One-loop contributions to the free energy}
\label{oneloops}
\end{figure}

Considering just the one-loop contribution the gauge-fixed Lagrangian is

\begin{equation}
{\cal L}_{QCD(R)} = \frac{1}{2} {\bar A}_{\mu}^{a} \left[ - \delta_{\mu \nu} D^2 (a) \right] {\bar A}_{\nu}^{a} + {\bar C}^a \left[ - D^2 (a) \right] C^a + {\bar \psi} \left( \Dsl_R + M - \gamma_0 {\cal M} \right) \psi ,
\end{equation}

\noindent where $C$ and ${\bar C}$ are the complex grassman ghost and antighost fields, respectively, which result from the gauge fixing. Then the path integral is

\begin{equation}
\begin{aligned}
Z_{QCD(R)} \left( \beta, \mu \right) &= \int {\cal D} A {\cal D} C {\cal D} {\bar C} \int {\cal D} \psi {\cal D} {\bar \psi} e^{- \int_0^{\beta} {\mathrm d} \tau \int {\mathrm d}^3 {\bf x} {\cal L}_{QCD(R)}}\\
&= \int {\cal D} ( g {\bar A}) {\cal D} C {\cal D} {\bar C} \int {\cal D} \psi {\cal D} {\bar \psi} e^{- \int_0^{\beta} {\mathrm d} \tau \int {\mathrm d}^3 {\bf x} {\cal L}_{QCD(R)}} .
\end{aligned}
\end{equation}

\noindent So the problem is reduced to performing Gaussian integrals. The final result is

\begin{equation}
Z_{QCD(R)} = \det \left[ \Dsl_R (a) + m - \gamma_0 \mu \right]^{N_f} {\det}^{-1} \left( - D_{Adj}^2 (a) \right) .
\end{equation}

The Yang-Mills theory result from the gluon and ghost contribution was calculated in \cite{Gross:1980br} and the result in terms of the Polyakov loop $P = {\rm diag} \{ e^{i v_1}, ... , e^{i v_N} \}$ is

\begin{equation}
\begin{aligned}
V_{YM} &= \frac{1}{\beta V_3} \ln \det \left( - D_{Adj}^2 \right)\\
&= - \frac{2}{\pi^2 \beta^4} \sum_{n=1}^{\infty} \frac{1}{n^4} \left[ \Tr_A \left( P^n \right) \right]\\
&= \frac{1}{\beta^4} \left[ \frac{1}{24 \pi^2} \sum_{i, j=1}^{N} [ v_i - v_j ]^2 \left( 2 \pi - [ v_i - v_j ] \right)^2 - \frac{\pi^2}{45} \left( N^2 - 1 \right) \right] ,
\end{aligned}
\end{equation}

\noindent where $[v] = v \mod 2 \pi$.

Now we can combine the boson and fermion contributions to get the full one-loop effective potential in terms of the Polyakov loop angles $v_i$. The result is

\begin{equation}
\begin{aligned}
&V_{1-loop} (P, m, \beta, \mu)\\
&= - \frac{1}{\beta V_3} \ln Z (P, m, \beta, \mu)\\
&= \frac{1}{\beta V_3} \left[ - 2 N_f \ln \det \left( - \left( D_R (P) - \mu \right)^2 + m^2 \right) + \ln \det \left( - D_{Adj}^{2} (P) \right) \right]\\
&=\frac{m^2 N_f}{\pi^2 \beta^2} \sum_{n=1}^{\infty} \frac{(\pm 1)^n}{n^2} \left[ e^{n \beta \mu} \Tr_R (P^{\dagger n}) + e^{- n \beta \mu} \Tr_R (P^n) \right] K_2 ( n \beta m ) - \frac{2}{\pi^2 \beta^4} \sum_{n=1}^{\infty} \frac{1}{n^4} \Tr_A ( P^n )\\
&= \frac{m^4 N_f}{3 \pi^2} \int_1^{\infty} {\mathrm d}t (t^2 - 1)^{3/2} \left[ g_{R,\pm} \left( \beta, m t, \mu, v \right) + g_{R,\pm}^{\dagger} \left( \beta, m t, - \mu, v \right) \right]\\
&+ \frac{1}{\beta^4} \left[ \frac{1}{24 \pi^2} \sum_{i, j=1}^{N} [ v_i - v_j ]^2 \left( 2 \pi - [ v_i - v_j ] \right)^2 - \frac{\pi^2}{45} \left( N^2 - 1 \right) \right] .
\end{aligned}
\end{equation}

If we take $\mu = 0$ then the result simplifies and we have

\begin{equation}
\begin{aligned}
&V_{1-loop} (P, m, \beta)\\
&=\frac{2 m^2 N_f}{\pi^2 \beta^2} \sum_{n=1}^{\infty} \frac{(\pm 1)^n}{n^2} {\rm Re} \left[ \Tr_R (P^{n}) \right] K_2 ( n \beta m ) - \frac{2}{\pi^2 \beta^4} \sum_{n=1}^{\infty} \frac{1}{n^4} \Tr_A ( P^n )\\
&= \frac{2 m^4 N_f}{3 \pi^2} \int_1^{\infty} {\mathrm d}t (t^2 - 1)^{3/2} {\rm Re} \left[ g_{R,\pm} \left( \beta, m t, 0, v \right) \right]\\
&+ \frac{1}{\beta^4} \left[ \frac{1}{24 \pi^2} \sum_{i, j=1}^{N} [ v_i - v_j ]^2 \left( 2 \pi - [ v_i - v_j ] \right)^2 - \frac{\pi^2}{45} \left( N^2 - 1 \right) \right] .
\end{aligned}
\end{equation}

\end{section}
\begin{section}{\label{reps}Higher dimensional representations}

In this appendix we show how to get higher dimensional representations of $\Tr_R P$ in terms of the fundamental $\Tr_F P$ where $P \in SU(N)$. In what follows we will refer often to Young tableau. A representation $R$ is referred to in terms of its Young tableau by a comma separated list enclosed in parentheses $(y_1, y_2, ..., y_{N-1})$. Thus, an arbitrary representation $R$, with a Young tableau as shown in Figure \ref{youngtab}, is represented by $(y_1, y_2, ... , y_{N-1}) = (p, q, r, ...)$.

\begin{figure}[h]
\begin{center}
\includegraphics[width=14cm]{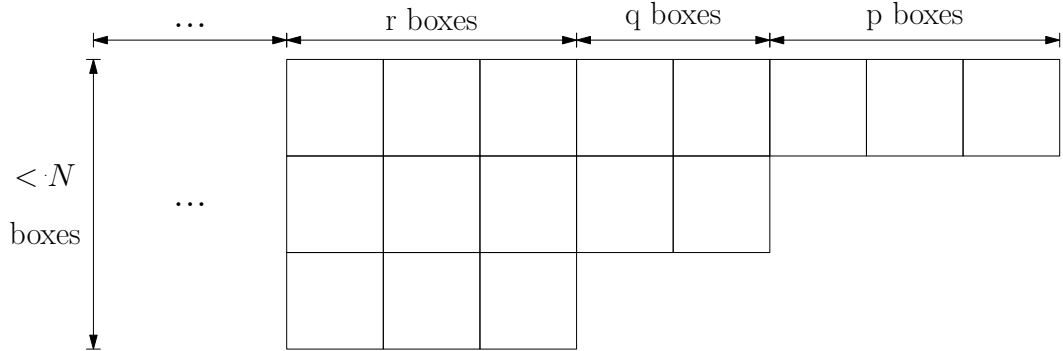}
\end{center}
\caption{Young tableau of an arbitrary representation $R$}
\label{youngtab}
\end{figure}

In what follows we will deal with representations that have either one or two rows of boxes in their Young tableaux, and so we will use the notation $(p,q)$ for Young tableaux that have $p$ columns with one row of boxes, and $q$ columns with two rows of boxes. However, the procedure is valid in general for all representations.

Higher dimensional representations $R$, with Young tableau $(p, q)$, of $SU(N)$ can be put in terms of the fundamental $F$, with Young tableau $(N, 0)$, by using the Frobenius formula \cite{Gross:1993yt,Gross:1998gk,Dumitru:2003hp}. In terms of the Polyakov loop the trace has the form

\begin{equation}
\Tr_R P = \frac{1}{n!} \sum_{{\bf j} \in S_n} \chi_R ({\bf j}) \left( \Tr_F P \right)^{j_1} \left( \Tr_F P^2 \right)^{j_2} ... \left( \Tr_F P^n \right)^{j_n} ,
\end{equation}

\noindent where $n$ is the number of boxes in the Young tableau of the representation $R$, the sum is over all $n!$ permutations ${\bf j} = \{ j_1, j_2, ..., j_n \}$ of the symmetric group $S_n$, and $\chi_R ({\bf j})$ is the group character, in the representation $R$, of the permutation ${\bf j}$ of the symmetric group $S_n$. These have a simple form when $R$ is a symmetric representation with Young tableau (n,0). This is given by \cite{Georgi:1982jb}

\begin{equation}
\chi_S ({\bf j}) = \frac{n!}{\prod_{k=1}^{n} k^{j_k} j_k !}.
\end{equation}

\noindent So for symmetric representations $R$ with Young tableau $(n,0)$

\begin{equation}
\Tr_{(n,0)} P = \sum_{{\bf j} \in S_n} \frac{1}{\prod_{k=1}^{n} k^{j_k} j_k !} \left( \Tr_F P \right)^{j_1} \left( \Tr_F P^2 \right)^{j_2} ... \left( \Tr_F P^n \right)^{j_n} .
\label{symm_frobenius}
\end{equation}

After we have the symmetric representations it is simple to get the rest by evaluating tensor products using Young tableau. Alternatively, all the representations can be found from the Frobenius formula, but the characters in other representations are not as simple.

Our procedure is best understood through example. Let us start with symmetric representations with Young tableau $(n,0)$. First we want to find all $n!$ permutations ${\bf j}$ of the symmetric group $S_n$.  We can obtain these from the elements of the symmetric group, $S_n$. The elements of $S_n$ for $n = 2, 3, 4$ are

\begin{equation}
\begin{array}{l|llllll}
S_2 : & (1)(2) & (1 2)\\
\hline
S_3 : & (1)(2)(3) & (1)(2 3) & (3)(1 2) & (2)(1 3) & (1 2 3) & (1 3 2)\\
\hline
S_4 : & (1)(2)(3)(4) & (1)(2 3 4) & (1)(2 4 3) & (2)(1 3 4) & (2)(1 4 3) & (3)(1 2 4)\\
& (3)(1 4 2) & (4)(1 2 3) & (4)(1 3 2) & (1 2)(3 4) & (1 3)(2 4) & (1 4)(2 3)\\
& (1)(2)(3 4) & (1)(3)(2 4) & (1)(4)(2 3) & (2)(3)(1 4) & (2)(4)(1 3) & (3)(4)(1 2)\\
& (1 2 3 4) & (1 2 4 3) & (1 3 2 4) & (1 3 4 2) & (1 4 2 3) & (1 4 3 2)
\end{array}
\end{equation}

\noindent Therefore the permutations ${\bf j}$ of $S_n$ for $n = 2, 3, 4$ are

\begin{equation}
\begin{array}{l|l|lllll}
n=2 & \{ j_1, j_2 \} & \{ 2, 0\} & \{ 0, 1 \}\\
\hline
n=3 & \{ j_1, j_2, j_3 \} & \{ 3, 0, 0 \} & \{ 1, 1, 0 \} & \{ 0, 0, 1 \}\\
\hline
n=4 & \{j_1, j_2, j_3, j_4 \} & \{ 4, 0, 0, 0 \} & \{ 1, 0, 1, 0 \} & \{ 0, 2, 0, 0 \} & \{ 2, 1, 0, 0 \} & \{ 0, 0, 0, 4 \}
\end{array}
\end{equation}

\noindent The permutations ${\bf j}$ can more simply be thought of as all possible solution vectors ${\bf j} = \{ j_1, j_2, ... , j_n \}$ to

\begin{equation}
1 j_1 + 2 j_2 + ... + n j_n = n
\end{equation}

\noindent with $j_i \ge 0, \, \forall \, i$.

At this point it is simple to get the higher dimensional representations $\Tr_R P$ from eq. (\ref{symm_frobenius}) where $R$ is a symmetric representation $(n,0)$. These are

\begin{equation}
\begin{aligned}
&\Tr_{(2,0)} P = \frac{1}{2} \left[ \left( \Tr_F P \right)^2 + \left( \Tr_F P^2 \right) \right] ,\\
&\Tr_{(3,0)} P = \frac{1}{3!} \left[ \left( \Tr_F P \right)^3 + 3 \left( \Tr_F P \right) \left( \Tr_F P^2 \right) + 2 \left( \Tr_F P^3 \right) \right] ,\\
&\Tr_{(4,0)} P = \\
&\hspace{4mm} \frac{1}{4!} \left[ \left( \Tr_F P \right)^4 + 6 \left( \Tr_F P \right)^2 \left( \Tr_F P^2 \right) + 3 \left( \Tr_F P^2 \right)^2 + 8 \left( \Tr_F P \right) \left( \Tr_F P^3 \right) + 6 \left( \Tr_F P^4 \right) \right] .\\ \vspace{1mm}
\end{aligned}
\label{symm-reps}
\end{equation}

Now that we have a procedure for obtaining all the $\Tr_R P$ for symmetric representations, we can get the rest by means of tensor products. Some useful tensor product identities (which can be obtained from Young tableau) are

\begin{equation}
\begin{aligned}
&(1,...,0) \otimes (1,...,0) = (2,...,0) \oplus (0,1,...)\\
&(1,...,0) \otimes (0,...,1) = (1,...,1) \oplus (0,...,0)\\
&(2,...,0) \otimes (0,...,1) = (2,...,1) \oplus (1,...,0)\\
&(2,...,0) \otimes (0,...,2) = (2,...,2) \oplus (1,...,1) \oplus (0,...,0)\\
&\left[ "..." \, \text{indicates the appropriate number of zeroes.} \right]
\end{aligned}
\label{tensor-prods}
\end{equation}

\noindent where we have brought back in the extra elements of $(y_1, y_2, ..., y_{N-1})$ for the purposes of getting the adjoint (A) representation $(1,...,1)$. Also, other representations of interest are the symmetric (S) $(2,...,0)$ and the antisymmetric (AS) $(0,1,...)$.

The first line of eq. (\ref{tensor-prods}) allows us to get the antisymmetric representation from the symmetric, which we found in the first line of eq. (\ref{symm-reps}), $\Tr_S P = \frac{1}{2} \left[ \left( \Tr_F P \right)^2 + \left( \Tr_F P^2 \right) \right]$. From the first line of eq. (\ref{tensor-prods})

\begin{equation}
\begin{aligned}
\Tr_{(0,1,...)} P &= \left( \Tr_{(1,...,0)} P \right)^2 - \Tr_{(2,...,0)} P\\
\Tr_{AS} P &= \left( \Tr_F P \right)^2 - \Tr_S P\\
\Tr_{AS} P &= \frac{1}{2} \left[ \left( \Tr_F P \right)^2 - \left( \Tr_F P^2 \right) \right] .
\end{aligned}
\label{astrace}
\end{equation}

The second line of eq. (\ref{tensor-prods}) allows us to get the adjoint representation from the fundamental (F) $(1,...,0)$ and antifundamental (${\bar F}$) $(0,...,1)$ representations

\begin{equation}
\begin{aligned}
\Tr_{(1,...,1)} P &= \left( \Tr_{(1,...,0)} P \right) \left( \Tr_{(0,...,1)} P \right) - 1\\
\Tr_A P &= \left| \Tr_F P \right|^2 - 1 .
\end{aligned}
\label{adjtrace}
\end{equation}

Other representations which are sometimes used are represented by $(2,1,...)$ and $(2,2,...)$. These are obtained from the third, and fourth line of eq. (\ref{tensor-prods}), respectively,

\begin{equation}
\begin{aligned}
\Tr_{(2,...,1)} P &= \left( \Tr_{(2,...,0)} P \right) \left( \Tr_{(0,...,1)} P \right) - \Tr_{(1,...,0)} P\\
&= \frac{1}{2} \left( \Tr_F P^{\dagger} \right) \left[ \left( \Tr_F P \right)^2 + \left( \Tr_F P^2 \right) \right] - \Tr_F P ,
\end{aligned}
\end{equation}

\begin{equation}
\begin{aligned}
\Tr_{(2,...,2)} P &= \left( \Tr_{(2,...,0)} P \right) \left( \Tr_{(0,...,2)} P \right) - \Tr_{(1,...,1)} P - 1\\
&= \frac{1}{4} \left[ \left( \Tr_F P \right)^2 + \left( \Tr_F P^2 \right) \right] \left[ \left( \Tr_F P^{\dagger} \right)^2 + \left( \Tr_F P^{\dagger 2} \right) \right] - \left| \Tr_F P \right|^2 .
\end{aligned}
\end{equation}

These representations are often named by their dimension $D$ which can easily be obtained using the "factors over hooks" rule $D = F/H$ \cite{Georgi:1982jb}. For $SU(3)$ and $SU(4)$ the above representations have the following dimensions.

\begin{equation}
\begin{array}{l|l|l|l}
\text{name} & (y_1,y_2,...,y_{N-1}) & D_{SU(3)} & D_{SU(4))}\\
\hline
\text{fundamental} \, ({\rm F}) & (1,...,0) & {\bf 3} & {\bf 4}\\
\text{antifundamental} \, ({\rm {\bar F}}) & (0,...,1) & {\bf {\bar 3}} & {\bf {\bar 4}}\\
\text{adjoint} \, ({\rm A}) & (1,...,1) & {\bf 8} & {\bf 15}\\
\text{symmetric} \, ({\rm S}) & (2,...,0) & {\bf 6} & {\bf 10}\\
\text{antisymmetric} \, ({\rm AS}) & (0,1,...) & {\bf {\bar 3}} & {\bf 6}\\
? & (0,...,2) & {\bf {\bar 6}} & {\bf {\bar {10}}}\\
? & (2,...,1) & {\bf 15} & {\bf 36}\\
? & (2,...,2) & {\bf 27} & {\bf 84}\\
\end{array}
\end{equation}

\noindent For these representations the notation in the case of $SU(2)$ needs to be considered specially since $(y_1 , ... , y_{N-1}) = (y_1)$ is usually not sufficient to describe Young Tableaux for representations with $y_i \ne 0$ where $i > 1$. Instead, we can use $(y_1 + y_1^*)$ to count the columns of boxes in a single row where $y_1^*$ is the number corresponding to $y_{N-1}$ above. This makes use of the fact that the fundamental and antifundamental representations are the same in $SU(2)$. Hence, the adjoint representation is the same as the symmetric. Also, in $SU(2)$ $y_2$ represents only singlet contributions so the antisymmetric representation becomes the singlet.

\end{section}


\end{document}